\documentstyle[epsf,12pt]{article}
\begin{document}
\begin{center}
\Large{\bf Searching for Quark-Gluon Plasma (QGP) Bubble Effects at
RHIC/LHC}\\
\large{S.J. Lindenbaum$^{a,b}$, R.S. Longacre$^a$, and M. Kramer$^b$\\
$^a$Brookhaven National Laboratory, Upton, NY 11973, USA\\
$^b$City College of New York, NY  10031, USA\footnote{This research was 
supported by the U.S. Department of Energy under Contract No. 
DE-AC02-98CH10886 and the City College of New York Physics Department}}
\end{center}
 
\begin{abstract}
Since the early eighties, we have shared with Leon Van Hove the 
following view. That if a QGP were produced in high energy heavy 
ion colliders, that its hadronization products would likely come 
from small localized in phase space bubbles of plasma. We develop 
a model based on HIJING, to which we added a ring of adjoining 
multiple bubbles in the central rapidity region. Our simulations 
were designed to be tested by the forthcoming RHIC STAR detector 
data for 65GeV/n $Au$ colliding with 65 GeV/n $Au$. We took into 
account background and resonance effects to allow a direct comparison 
with the data.  Later 100 GeV/n $Au$ colliding with 100 GeV/n $Au$ 
and LHC data could also test these ideas. We used two charged particle 
correlation's as a sensitive method to test for bubbles.  
\end{abstract}
 
\section{Introduction} 

If Quantum chromodynamics (QCD) is correct there is no doubt that 
under conditions that exist in Lattice Gauge Theory (LGT) calculations, 
a large volume quark-gluon plasma (QGP) is expected to be created. 
However one can question whether high energy heavy ion collisions in 
the RHIC collider reproduce the LGT calculation conditions well 
enough, that production of a detectable large volume of QGP occurs.  
This has more or less been assumed in many theoretical calculations. 
However, to our knowledge no-one has shown that the actual dynamics
existing at RHIC would allow this to occur with detectable 
cross-sections. Two Lorentz contracted heavy ions pass through each 
other in the short times available.  There are turbulent ever changing 
dynamics of the environment resulting from their interaction. This 
does not give one assurance that even the very basic LGT requirements 
of thermal and chemical equilibrium  can be met.  Certainly this 
has not been demonstrated by any theoretical work we are aware of. 
Therefore from the early eighties onward, as documented in the first 
four RHIC experimental workshops from 85-90\cite{workshop}, we have 
concluded this is not likely to happen.
  
Instead our view has been that if QGP is created in a high energy heavy 
ion collider such as RHIC or LHC, it is more probable that local 
fluctuations would produce one to many droplets (clusters, chunks or 
bubbles) of QGP. These would possibly be detectable, especially if 
they were localized in phase space. Leon van Hove had this view, and 
did string model calculations\cite{van83,van87}, which resulted in small 
droplets of QGP being formed by the breaking of stretched strings. 
These QGP bubbles were localized in rapidity, and gave rise to rapidity 
bumps or peaks in the rapidity or psuedo-rapidity distribution.
 
We have previously published\cite{lind00} a treatment of the single
(spherical-no longitudinal expansion) bubble case (similar to the 
Van Hove type), which serves as an introductory paper for the general 
subject. It is possible that with enough statistics one could in 
principle find single to a few bubble events. We have concluded that 
multiple bubble formation (which we mentioned but did not do calculations 
for in Ref. \cite{lind00}) is the more probable general case. Therefore 
this case should be treated for realistic attainable statistics. Of
course a large number of bubbles in the multiple bubble case will 
obscure the resolution and observation of single bubble phenomena 
such as rapidity bumps etc. Furthermore the overall result would 
likely appear similar to a thermal model. However, this likely occurs 
because particles from different parts of space go into the same phase 
space. Assume all the bubbles were Van Hove spherical bubbles at rest 
(i.e. located at mid rapidity). Even though they are at different 
space points, they would add up to be equivalent to one big Van Hove 
spherical bubble. If we now give motion to each bubble along the 
beam direction, they will smear out in momentum space. Therefore 
to detect the effects of the multiple bubble case, we must carefully 
choose a region where phase space focusing will allow the addition 
of multiple bubble effects.  Thus we must find a part of momentum 
space that is highly 
correlated to position space. If we could force bubbles to remain 
at rest, then their particles would add together forming a rapidity 
bump at $\eta=0$. Even if this happened it would be difficult to get 
a clean signal over background in this region, because background 
particles from soft fragmentation end up at central rapidity. Thus 
it is important to find a phase space region free of such background. 

At RHIC the pre-hadronic matter is being pushed in the transverse 
direction\cite{adler1} building up transverse momentum $p_t$.  
Bubbles that are pushed along with this flow will hadronize into 
particles focused over a limited range of angles. In this paper we 
will model such bubbles, and address backgrounds which mimic bubble 
effects. These are mainly jets and to some extent resonance's. 

The above states the goals of this paper. We will treat the case 
of an approximately maximum number of multiple bubbles in one outer 
ring around the blast region. These bubbles contribute to the final 
hadronization of particles coming from the QGP.  This may not occur in 
the actual case. Therefore only future data analysis, can shed 
light on this.  We will find that a reasonable theoretical treatment can 
lead to methods for detecting QGP bubbles.
 
In this paper we are essentially limiting ourselves to simulations and
analysis suitable for comparison with forthcoming RHIC data.  We will 
utilize relevant parts of the considerable body of data that has 
already been published. It is expected that in the near future  
data from RHIC for 65GeV/n $Au$ colliding with 65 GeV/n 
$Au$ could test these ideas. However, similar methods could be applied 
to higher energies at RHIC and LHC.
 
\section{General Considerations}

At RHIC the pre-hadronic matter is being pushed outward in the transverse
direction\cite{adler1}. Particles with higher transverse momentum
($p_t$) are pushed more than particles with lower $p_t$. Analysis of 
pions by Hanbury Brown-Twiss (HBT)\cite{adler2} methods have shown 
that at low $p_t$ the source size is about 6 fm, while above 0.8 GeV/c 
the source size becomes about 2 fm.  This allows phase space focusing 
to form a reasonable signal in the 2 fm source size region.  These
measurements imply that the viewed region of the initial position 
space becomes smaller as one selects higher $p_t$ particles. Pions at
a $p_t$ $\sim$ 1-2 GeV/c will be coming from the outer regions of           
the expanding fireball in regions where the HBT original source size radial
width is of the order of 2 fm.  Softer pions will mainly have a 
radius of within about 6 fm. This supports a rough estimate of $\sim$ 
6+2 fm to be the transverse radius of the fireball that is emitting 
hadronization particles. One should keep in mind that this is a 
quantum mechanical system with dynamical and turbulent changes. 
Therefore the previous arithmetic, and subsequent arithmetic is 
to be considered in the sense of very crude estimates. 

One should note that particles above 2 GeV/c will likely have jets 
as a sizeable source of contamination.  Thus we believe we can work 
with a window in $p_t$ of 0.8 to 2 GeV/c, to search for multiple 
bubble effects. We chose the upper end of the window to avoid the 
hard physics region above 2 GeV/c, which would increase the background 
for the effects we are looking for. The lower end of the range is 
chosen to maintain the space momentum correlation of our signals, 
and therefore enhance them.
 
Sometimes we vary the window somewhat to investigate a particular point.
From Ref. \cite{lind00} we found that the single plasma bubbles have 
a mean $p_t$ of about 0.5 GeV/c. Thus the azimuthal angular range in 
$\Delta\Phi$ 
can be crudely estimated for the bubbles in our window.  With an 
average $p_t$ of about 1 GeV/c, and  the above right angle momentum, 
we form an $\sim$ 30$^circ$ angle. Thus we can assume an approximation 
that spherical bubbles have an angular range in $\Delta\Phi$ of about 
30 degrees.  This however is very similar to the angular spread 
of jet fragmentation, making it  difficult to separate the spherical 
bubbles $\Delta\Phi$ distribution from the $\Delta\Phi$ distribution 
of jet fragmentation. It might be noted that there are 
arguments for jet quenching\cite{wang91} which would improve our signal 
compared to the assumed background. However we will ignore jet 
quenching in our simulations in order to be very conservative in 
drawing our conclusions.
 
We know there is a longitudinal expansion-the Landau fireball 
effect. The value of this longitudinal expansion has to be 
determined from analyzing the data.  However the Landau  
longitudinal expansion of our bubbles used in our previous 
paper\cite{lind00} was probably too large. Therefore we will choose a 
reasonable value intermediate between that and the value for a spherical 
bubble (which has zero longitudinal expansion) for our simulations.
 
Figure 1 contains a sketch of the assumed bubble geometry, and details 
of how bubbles are embedded are given in the caption. In the 
language of Van Hove, the string stopping after breaking is not 
complete, so that longitudinal expansion is left 
in the strings or bubbles. The longitudinal expansion will increase 
the angular spread of $\Delta\eta$ due to bubbles, which will distinguish 
bubbles from jet fragmentation. However it will spread out the bubble 
signal, but this may not matter.  Increasing the energy in the bubble,
which also reasonably could occur, would enhance the bubble signal. This 
leads to our using an about 50$^\circ$ range for the bubbles in the 
psuedo-rapidity direction for our simulations. The angular range of 
the bubbles in the azimuthal direction is about 30$^\circ$, as 
previously discussed.

From HBT work, previously referred to, we can estimate the bubble 
would have a radius of about 2 fm, and thus has a diameter of about 
4 fm.  The rapid transverse expansion in the blast region pushes 
high density pre-hadronic matter from the central regions outward to 
where it hadronizes\cite{adler2}. We assume a single outer ring of 
bubbles at the outer circumference of the blast region would be filled 
with bubbles at hadronization.  The bubbles provide the hadronization 
coming from the QGP. An angular range of 30$^\circ$ is one twelfth of 
360$^\circ$. Since each bubble has a diameter of about 4 fm, the 
circumference to cover the entire azimuth, would be approximately 48 fm.
A circumference of 48 fm implies a radius of about 8 fm for the ring of 
bubbles. The number of 8 fm is consistent with our HBT picture 
presented previously. Inside this outward shell of bubbles there 
may be other overlapping bubbles, but the ring we have chosen will 
predominantly contribute to the mid rapidity region.
 
Using the work in Ref. \cite{dumi01} that bubbles of a 2 fm radius (one 
of the choices) bubble size for a RHIC event would have about 40 domains
(which we call bubbles) with energies of about 3 GeV per bubble going 
into charged pions. We calculated the energy per domain (bubble) using 
the information in their paper.  For our simulations we felt it was 
reasonable to use the 2 fm bubble size in employing the methods  
in \cite{dumi01} since it was  the choice we arrived at from the 
HBT work.  We also decided that in this first simulation it was 
reasonable to use our calculation (based on\cite{dumi01}) of about 3 
GeV per bubble hadronizing into charged pions, in order to avoid 
being arbitrary. Obviously this value has to be considered a parameter 
which could be determined in conjunction with data analysis when it 
becomes available. However it should be noted we are not using any 
of the other detailed work in Ref. \cite{dumi01}. 

We used an average of 13 bubbles in a ring at approximately 
mid-rapidity in each central 65 GeV/n $Au$ on 65GeV/n $Au$ event. 
This  fills the ring of bubbles, whose contributions would dominate 
what is seen at central rapidities at the RHIC STAR Detector. This 
detector would be the most likely near future source of experimental 
results to check these ideas. We used an average of 3.25 GeV per 
bubble. The energy was increased from 3.0 GeV per bubble, since we 
are producing more than pions going into charged particles. This led 
to an average of 1.95 charged particles going into the cuts we will use. 
 
In our bubble scenario each QGP bubble is an uncharged, gluon dominated, 
color singlet system.  Thus when the bubble hadronizes the total 
charge of the particles coming from the bubble is zero. Since we 
are selecting a $p_t$ range where we expect the bubble concentration 
to be rich, we should therefore see a suppression of charge 
fluctuations.   This occurs because the charge fluctuations coming 
from a localized QGP bubble should be less by a factor $\sim$4, 
than charge fluctuations coming from an ideal pion gas. This 
subject will be addressed in the next section.

\section{Charge Suppression Effects Due to QGP Formation}
 
We now address recent papers on charge fluctuation suppression 
calculations, of the ratio of positively charged and negatively 
charged pions as a signal for QGP formation. In the letters of 
S.~Jeon, and V.~Koch\cite{jeon}, and M.~Asakawa, U.~Heinz, and 
B.~Mueller\cite{asakawa}, they concluded that a parameter 
$D = {4 < \Delta Q^2>\over <N_{ch}>}$ for mesons  evaluated for event 
by event charge fluctuations, is $\sim$4 for an ideal pion gas not 
produced by QGP. They estimate it would be approximately 1 for pions 
originating from a QGP both from LGT or ideal gas calculations. 
Correcting for resonance effects in actual observations, they 
concluded the observed $D$ would become 3-2 respectively. Therefore 
they argued observing these reduced charge fluctuations would serve 
as a distinct signal for QGP production. We pointed out\cite{lind01} 
that their method of treatment\cite{jeon,asakawa} allowed color charge 
fluctuations, and kinematic mixing effects, which they overlooked. 
These color charge fluctuations, and kinematic mixing effects were 
shown by us to be important, and model dependent. They could modify 
the result by washing out or even entirely eliminating the charge 
suppression effects predicted\cite{lind01}. In regard to the kinematic 
mixing effects, it should be noted that in their treatments, even if 
locally in space one has a charge fluctuation suppression of $\sim$ 
70\% to 50\%, kinematic mixing of position space into momentum space 
causes electric charge fluctuations to increase over a wide region,
thus reducing suppression. Cuts on the particles one measures will also 
reduce the calculated suppression. If one gains particles from other parts 
of position space, that reduces the calculated suppression. Thus 
we concluded that it is unlikely their arguments could support their 
predictions for observation of electric charge fluctuation suppression, 
even if large volumes of QGP were produced as they assumed.

\section{Electric Charge Fluctuation Suppression in the Bubble
Scenario}
 
In our bubble scenario each QGP bubble is a very localized color
singlet and uncharged system. Thus when the bubble hadronizes 
the total charge of the particles coming from the bubble has 
to be approximately zero. Therefore due to the localization, 
color charge fluctuations, and most of the kinematic mixing 
effects which can drastically change the suppression predicted 
in \cite{jeon,asakawa}, are negligible. Since we are selecting a 
$p_t$ range where we expect the bubble effects to be substantial, 
we should see a suppression of charge fluctuations in the charged 
particles coming from the QGP bubbles . The large predicted 
reduction by a factor of $\sim$4 of charge fluctuations which is 
reduced to $\sim$2-3 by resonance's in\cite{jeon,asakawa}, will not be 
achieved by our measurements because we take account
of the presence of background particles, and the loss of some
of the plasma particles out of our cuts. However there will be 
predicted a measurable reduction, and it might be considerable.
 
In a study of resonance effects, we will find that conclusions 
from our simulations for our chosen signals are not significantly 
affected by resonance background. Thus we expect under the 
bubble scenario that we are following, there will be predicted 
observable charge suppression. This will be good evidence for 
QGP formation if observed. One needs to await analysis and 
availability of the relevant data to test, and if it appears 
relevant, to optimize these ideas. 
 
\section{HIJING Based Model}

Our first objective is to construct a model which will hopefully take
account of the most important effects due to QGP bubbles, background,
and resonance effects etc. This would allow a direct comparison of the 
model predictions and the future RHIC data  in a reasonably quantitative
manner. We now make a model based on the HIJING event 
generator\cite{wang91}. For $Au$ on $Au$ at 65 GeV/c per nucleon. 
HIJING is a good choice to base a simulation on since  
HIJING has been successfully used to fit, and help with the analysis 
of RHIC data in numerous instances. However HIJING has an important 
missing part for our $Au$ on $Au$ simulation. That is elliptic 
flow which has been measured at RHIC\cite{ack01}. Not taking account 
of elliptic flow would unrealistically modify our simulation results. 
Therefore we have modified HIJING to include relevant elliptic flow 
effects. HIJING has two relevant sources of particle production: 
Jets which fragment into particles which are referred to as jet 
particles, and the soft particles which come from beam jet fragmentation. 
The jet particles are not flat in azimuth but bunch around the 
jet axis. The beam jets fragmentation's are very flat in azimuth. 
 
To take account of the observed elliptic flow we modify the 
distribution in azimuth of the soft particles (beam jets) so that 
we develop a $\cos 2\Phi$ component about a fixed axis for each 
$Au$ on $Au$ simulated event. Into each central $Au$ on $Au$ event we 
have added ,on the average, 13 adjoining bubbles in a single ring in
the central rapidity region  to replace the mini-jets.  We are 
assuming in essence that region is perhaps the source of bubble 
production. Each bubble contributes from 1 to 4 charged particles to 
the $\eta$ range of +0.75 to -0.75, with $p_t$ greater than 0.8 GeV/c 
and less than 2.0 GeV/c. This $p_t$ cut has its lower bound chosen to 
maintain the space momentum correlation which enhances our signal, 
while the upper bound is chosen to avoid contamination from the hard 
physics region above 2 Gev/c, and it is our most relevant $p_t$ cut.
However some of the time we use 1.2 GeV/c for the lower bound  to
investigate and separate various effects as indicated on some 
figures and in the text.  

The total charge of each bubble was set to zero, which is 
appropriate for a QGP bubble. We then generated 100,000 bubble
events with impact parameter ranging from 0.0 to 4.0  fm. 
We also generated 100,000 events of our modified HIJING
taking into account relevant elliptic flow effects.
 
From this point onward, we  have made our assumptions on bubble              
geometry, and how to embed them.  Therefore our efforts in the 
remainder of the paper are devoted to how to separate background and 
resonance effects etc. Thus allowing us to detect the effects of the 
bubbles in the RHIC experiment cited.
 
In Fig. 2 we show the $\Delta\Phi$ correlation generated by the above  
simulations, including the embedded bubbles and relevant elliptic flow in 
our modified HIJING. For comparison we make use of our modified HIJING
which has the beam jets modified by elliptic flow, and the expected 
mini-jets predicted by HIJING.
 
For $Au$ on $Au$ with an energy of 65 GeV per nucleon we used the 
standard 2 GeV/c QCD cutoff. One obtains an average of 17.6 jets per 
event from which 13.3 charged particles fall into our cuts (Fig. 2). 
It appears the STAR detector at RHIC is the best bet for experimentally 
checking in detail the theoretical calculations in this paper. 
Therefore we have added two track merging that one sees in the STAR 
detector\cite{adler2}.  This inefficiency only effects small 
$\Delta\Phi$'s. In Fig. 2, we see that the plasma bubbles have a 
stronger correlation than the standard mini-jets. In order to make 
quantitative comparisons for angles less than 60$^\circ$ where we 
expect the bubbles to contribute, we calculate $\chi^2$ in all angular 
correlation calculations with this cut. In Fig. 2 the $\chi^2$ is 53 
for 8 bins (Degrees of Freedom). This is an order of $\sim$8$\sigma$ 
effect. The 
theory predicting the number of mini-jets is not exact, and we wish 
to be conservative. Therefore we ask the question, by how much of 
a factor do we have to increase the HIJING predicted jets to make 
the bubble effect difference in $\Delta\Phi$ become indistinguishable 
from HIJING with the added jets. 
 
In Fig. 3 we show that from an 100,000 event simulation, that 
arbitrarily doubling the number of mini-jets in HIJING causes the 
$\chi^2$ to drop to 6 resulting in no difference between bubbles, and
arbitrarily increasing the number of mini-jets generated by HIJING to
double. This represents a very conservative, and probably an 
excessively overdone approach, especially since reasonable arguments 
exist that actually jet quenching rather than enhancement 
occurs\cite{wang91}. The agreement between HIJING plus bubbles and
HIJING plus double jets is also good if we choose a tighter cut with
a $p_t$ range 1.2 to 2.0 GeV/c (see Fig. 4, for which $\chi^2=10$).
 
If we bin in $\Delta\eta$ and plot HIJING plus bubbles, and HIJING plus 
double jets, we see the correlation for the bubbles compared to the 
jets in Figs. 5-8 for various $\Delta\eta$ ranges. The $\chi^2$ for the 4 
Figs. are 1, 4, 14, 1 for 3DF. Only Fig. 7, shows a difference, 
with some statistical significance of over 3$\sigma$ for particles 
which are separated by $\Delta\eta$ of 1.05. The difference is that the 
correlation is wider for the bubbles compared to the jets. This difference 
in width along the $\eta$ direction, is expected from the Landau 
longitudinal expansion of the bubbles. With more statistics in the
simulation, we can expect that the differences in width along  
the $\eta$ direction would become more evident, and have better statistical
significance due to the nature of the effects longitudinal expansion
produces.
  
Let us look at an angular correlation of the angle 
between particles (opening angle $\cos\Theta$). Figure 9 indicates 
that the bubbles have a wider correlation than the double jets. The 
$\chi^2$ for Fig. 9 is 26 for 8 DF which is also $\sim 3\sigma$. Next 
let us do $\Delta\Phi$ correlations for like and unlike charges 
separately. In Fig. 10 we show this correlation for the $p_t$ range 
0.8 to 2.0 GeV/c. We see that the difference between the unlike and 
like charges $\Delta\Phi$ correlation is larger for the bubbles
than for double jets. The $\chi^2$ is 70 for DF=24 (3X8=24) which is 
a 5$\sigma$ effect. This difference is due to the zero charge of the 
bubbles while jets only have a reduced charge. This represents the charge 
fluctuation suppression effects we are looking for. We can form a 
measure of charge fluctuations by looking at the charge difference, 
event by event, for particles which lie in our $p_t$ range (0.8 to 2.0 
GeV/c), and $\eta$ range ($\vert\eta\vert <0.75$).  For the mean of 
the charge difference, with our average of 108 particles per event,
we get 4.0 positive charges per event. The width for the
double jets is 10.4 particles. The square root of 108 is 10.4. Thus 
the width for the double jets is consistent with a purely random 
charge fluctuation result. The width for the bubbles is 9.7 particles 
thus being consistent with 95\% of a random charge fluctuation result. 
When pairs of charges are created and go into the 
$p_t$ window (0.8 to 2.0 GeV/c) then the charge fluctuations are 
reduced. However when one charge goes into the window and one 
outside there is a  random addition.
 
Since we see a net positive charge this implies baryon transport 
to the central region. We are summing over impact parameters of 
0.0 to 4.0 fm, and thus have a varying fluctuation of baryon transport. 
For more central events the net positive charge would be larger, and 
for the less central events this positive charge would be smaller. 
This effect causes a larger than random charge fluctuation. It also 
appears that these effects cancel out for our HIJING simulation with 
the double jets. 
 
The bubble fluctuations appear somewhat smaller. This is because not all the 
bubble are contained in the above cuts. Pions from the bubble end up 
having $p_t$ near the lower edge of the $p_t$ range. The kaons are 
boosted to the mid-range, and protons are near the upper range. As 
stated previously. The upper end of the range is chosen to avoid the 
hard physics region above 2.0 GeV/c, and the lower end of the 
range is chosen to maintain the space momentum correlation. 

In the future the planned Time of flight system which surrounds the 
central TPC at RHIC is expected to become available. We will then be 
able to do much more detailed treatments similar to the above. We can
analyze $\pi^+$ and $\pi^-$, $K^+$ and $K^-$, and $p$ and $\bar p$, 
instead of just using, all positive and all negative charge pairs.
Eventually when there is enough experimental statistics available to
compare with, we can do multiple particle correlation's for larger 
numbers of particles.

\section{Estimating Resonance Effects with a Resonance Gas Model}
 
In order to estimate resonance effects, we use a model based 
on simple thermal ideas, plus resonance's expected in a hot hadronic
system. For thermal particle production we will use a simple 
factorized form for $p_t$ and $y$ of our particles and resonance's.
 
\begin{equation}
{d^2N\over{dp_tdy}}
=Ap_t{e^{-{M_t\over{Te}}}e^{-{y^2\over{2\sigma^2_y}}}}.
\end{equation}

The inputs are the mass of the particle or resonance ($M$)(GeV), 
the calculated transverse mass($M_t$)(GeV),and the thermal
temperature of the particle ($Te$)(GeV). A Gaussian width in rapidity 
$\sigma_y$ is also used.  The resonance mass is smeared by a 
Breit-Wigner form. A typical resonance is the $K^\ast$ which has a form:
 
\begin{eqnarray}
W(M)&=&{\Gamma^2_{(q)}M^2_{k^\ast}}\over
(M^2-M^2_{k^\ast})^2+{\Gamma^2_{(q)}M^2_{k^\ast}} \nonumber \\
\Gamma(q)&=&{{2\Gamma_{k^\ast}(q/q_0)^3}\over{1+(q/q_0)^2}}.
\end{eqnarray}
 
\noindent{where} $W(M)$ is the mass weighting of the $K^\ast$, and 
$\Gamma (q)$ is the total width (GeV). $M_{k^\ast}$ is the mass (GeV) of 
the $K^\ast$ and $q$(GeV/c) is the $\pi^- K$ C.M. momentum, with $q0$
(GeV/c) being the saturation momentum. The powers 3 and 2 are 
derived from $2l+1$ and $2l$, where $l$ is the angular momentum of 
the $\pi^- K$ system. We also need to add elliptic flow\cite{ack01} 
to our particle production. A very simple $p_t$ dependent $V2$ parameter
is used for our resonance gas as we used for the beam jets in HIJING. 
We used the form

\begin{equation}
E{d^3N\over{dp^3}}={1\over{2\pi p_t}}{d^2N\over{dp_tdy}}
[1+2V_2p_t\cos 2(\Phi-\Phi_R)]
\end{equation}
  
\begin{center}
\begin{tabular}{|c|r|r|r|r|}\hline
\multicolumn{5}{|c|}{Table I}\\ \hline
particle & number & temp & width & $V_2$ \\ \hline
$\pi^+$ & 115 & .250 & 3.00 & .12 \\ \hline
$\pi^-$ & 115 & .250 & 3.00 & .12 \\ \hline
$\rho^0$ & 350 & .290 & 2.90 & .12 \\ \hline
$\omega$ & 252 & .290 & 2.90 & .12 \\ \hline
$\eta$ & 505 & .290 & 2.90 & .12 \\ \hline
$k^+$ & 133 & .260 & 2.20 & .09 \\ \hline
$k^-$ & 116 & .260 & 2.20 & .09 \\ \hline
$k_s$ & 125 & .260 & 2.20 & .09 \\ \hline
$k^{\ast +}$ & 19 & .260 & 2.20 & .06 \\ \hline
$k^{\ast -}$ & 17 & .260 & 2.20 & .06 \\ \hline
$k^{\ast 0}$ & 35 & .260 & 2.20 & .06 \\ \hline
$p$ & 45 & .320 & 1.80 & .03 \\ \hline
$\bar p$ & 33 & .320 & 1.80 & .03 \\ \hline
$\Lambda$ & 71 & .360 & 1.80 & .03 \\ \hline
$\bar\Lambda$ & 50 & .360 & 1.80 & .03 \\ \hline
$\Sigma^+$ & 33 & .360 & 1.80 & .03 \\ \hline
$\Sigma^-$ & 33 & .360 & 1.80 & .03 \\ \hline
$\bar\Sigma^+$ & 21 & .360 & 1.80 & .03 \\ \hline        
$\bar\Sigma^-$ & 21 & .360 & 1.80 & .03 \\ \hline
$\Xi^0$ & 17 & .330 & 1.80 & .03 \\ \hline               
$\Xi^-$ & 17 & .330 & 1.80 & .03 \\ \hline
$\bar\Xi^0$ & 13 & .330 & 1.80 & .03 \\ \hline         
$\bar\Xi^+$ & 13 & .330 & 1.80 & .03 \\ \hline         
$\Omega^-$ & 4 & .300 & 1.80 & .03  \\ \hline          
$\Omega^+$ & 4 & .300 & 1.80 & .03  \\ \hline
\end{tabular}
\end{center}
 
Here there is only one parameter which is $V2$. We can also add 
jets from HIJING\cite{wang91} as we did above. In 
Refs. \cite{jeon,asakawa,lind01} one expects a thermal resonance gas 
system should have a suppression of charge fluctuation of the order 
of 20\% (if $D$ is $\sim 3$ due to resonance effects whereas it would 
be $\sim 4$ without resonance effects).  Using the resonance to pion 
ratio of Ref. \cite{herr} we can generate central 65 GeV/n  $Au$ on 65 
Gev/n $Au$ events that are very close to the HIJING simulation, using 
a Boltzman temperature of 0.180 GeV, and a Gaussian width rapidity of 2 
units. It is important to note that in order to make comparisons 
between the different models, we need to reproduce the single particle 
($p_t$) and psuedo-rapidity distributions.
 
Then for a pseudo-rapidity of $-0.75 < \eta < 0.75$, we can look at 
the distribution of net charge. From this net charge distribution we 
can determine the charge suppression. Keeping the final yield of 
particles the same, we increase the resonance's until we achieve a 20\% 
reduction in charge fluctuation which represents the resonance effect 
used in [9-11]. This is roughly consistent with observations. The net 
charge mean is 9.5 with a width of 20. For a random charge 
fluctuation system the width should be 25. Adding jets from HIJING 
to our resonance gas does not appreciably change the above results. 
In table 1, we give the number of particles and resonance's used and 
the temperatures and rapidity widths for each plus the $V2$ parameter.
Figure 11 shows the $\Delta\Phi$ comparison with resonance plus jets 
to HIJING plus bubbles. We have a $\chi^2$ of 22 for 8 $DF$ which is a 
$\sim 2.5 \sigma$ difference, and thus the two are statistically 
equivalent to being the same, since we consider at least $3\sigma$ 
required for minimal statistical significance. In Fig. 12 we make a
$p_t$ cut ($1.2 < p_t< 2.0 $GeV/c) and see that HIJING plus bubbles 
has a considerably increased correlation as a function of $\Delta\Phi$ 
in our usual 0-60$^\circ$ cut region. There is a $\chi^2$ of 
206 which corresponds to over 20$\sigma$.  We see that resonance's
 do not decay most of the time into particles with enough 
$p_t$ values to satisfy this cut. Thus by making a tighter $p_t$ 
cut the correlation in $\Delta\Phi$ due to resonance's is greatly 
reduced, and we can use this tightening of cuts to reveal the 
correlation in $\Delta\Phi$ without significant contributions from 
resonance effects.  We now can make a charge fluctuation analysis 
of the two models inside our standard cuts. The results for 
resonance plus jets is a 1\% reduction in the width of the net 
charge (10.05 compared to 10.15), whereas for HIJING plus bubbles 
we see a reduction of 5\% in the width (9.73 compared to 10.29). 
 
\section{Simulation of Bubbles with 3 Times the Energy per Bubble}
 
Up to this point in our model building, we have calculated the energy
in a bubble using the work in Ref. \cite{dumi01}, and replacing one of 
their domains with a bubble of about 2 fm radius, which is one of their 
choices they mention. Our reason for selecting a 2 fm bubble is based 
on using the previously mentioned HBT work on source size, and adapting 
it to our bubble scenario. That seems somewhat reasonable as a starting 
point, in order to avoid an arbitrary selection by us. However, one 
has to admit that the energy per bubble is really a parameter and that 
their value could be off by a considerable factor. If our use of 
calculated energy per bubble based on Ref. \cite{dumi01} estimates is 
too large, we feel it would become increasingly difficult as the 
energy per bubble decreased to observe multiple bubbles effects using 
the methods explored above.  There are many unknown uncertainties in 
the energy per bubble used from Ref. \cite{dumi01}, including the 
turbulent fluctuation phenomena which exist. Therefore it is 
reasonable to speculate what would happen if the bubble energies 
were bigger. Therefore we have rerun our bubble simulation, using 3 times
the energy per bubble, that was used originally. Figure 13 shows the
$\Delta\Phi$ correlation with the bigger energy bubbles compared 
to the regular modified HIJING and jets of Fig. 2.  This is a 
very large effect with a $\chi^2$ of 1840, which is well over 20 
$\sigma$. Next we show the unlike and like charge sign comparison in 
Fig. 14. Here the $\chi^2$ is 4399, thus we see again a very large
difference well over 20 $\sigma$ between like and unlike charge pairs. 
If we look at the charge difference distribution of the 3 times more 
energy per bubble simulation, we find a mean of 4 with a width of 7.6. 
A pure random charge fluctuation case has a width of 10.4 thus 
leading to about a 27\% reduction of charge fluctuation, with 3/4 
of the particles in our cuts coming from the plasma bubbles. 
 
\section{Most Extreme Pure Neutral Resonance Case} 
 
Finally, let us explore the most extreme totally unjustified
case, which should give a maximum charge suppression by generating 
a pure neutral resonance system. This possibility has never been
seen in heavy ions collisions. It violates isospin symmetry along 
with other reasonable considerations. We generated this case merely 
to demonstrate that our bubbles focus much more of their decay
particles into the cuts than a resonance case can. We will consider
charged particles generated by only resonance's decaying into the 
region of our standard cuts ($0.8 < p_t< 2.0-0.75<\eta< 0.75$). 
For our neutral resonance's we will use the $\rho$ and the 
$\sigma$ mesons. The $\rho$ is given by the $I=1$, P-wave $\pi\pi$
phase shifts, while the $\sigma$ is given by the $I=0$, S-wave 
$\pi\pi$ phase shift(see Ref. \cite{gray}). 
 
In the heavy ion final state, we expect many resonance's will be 
formed by pion re-scattering in the final state. Therefore the
$\sigma$ meson will have a mass shift due to re-scattering like the
$a1$ meson does in diffractive production\cite{aaron}. This 
re-scattering will create a threshold peak in the di-pion effective 
mass spectrum. We can show the unlike charge di-pion correlation as a 
function of the effective mass if we plot the ratio of the unlike charge  
pions over the like charge pions. In Fig. 15 we show this ratio 
versus the di-pion effective mass. There is a threshold bump given 
by the $\sigma$ and a second bump given by the $\rho$. In Fig. 16 we show
the $\Delta\Phi$ correlation that we have generated using our two
resonance's. Again we must make sure that both the single particle
$p_t$ and pseudo-rapidity spectra are the same as in our other 
simulations. The values used are shown in Table 2. In the table
we see that we need two different temperature sources of
neutral resonance's in order to obtain the $p_t$ spectrum. 
 
This time we compare our neutral resonance system with the 
resonance gas system plus bubbles. The $\chi^2$ for these two 
in Fig. 16 is 11 which is within $\sim 0.5\sigma$ the same. We 
have compared this extreme pure neutral resonance case to our standard
bubbles case where $1\over 4$ of the particles come from the bubbles, 
while 100\% of the particles come from the neutral resonance's. 
However again if we make a narrower $p_t$ cut ($1.2 < p_t< 2.0$) 
the bubble correlation is larger by a $\chi^2$ of 36 which is a 
$5\sigma$ effect (Fig. 17). Also the width in $\eta$ is different 
than that of the bubbles. This can be seen by looking at the 
opening angle correlation (Fig. 18) which should be compared to 
Fig. 9. The $\chi^2$ in Fig. 18 is 50 which is a $7\sigma$
effect. The reduction of charge fluctuation for the pure neutral 
resonance system is an interesting result. This system has a net 
charge of zero, and would have no charge fluctuation if there were 
no decays.  It has only a 6\% reduced fluctuation in our case. 
The cause of this effect is decay particles leaking out of the cuts. 
This 6\% is the total amount possible from neutral resonance's
decaying into our cuts even in this extreme unrealistic case.
 If we increase our bubble energy by 3 times we achieve a 27\% 
reduction in charge fluctuations. For the resonance gas plus bubbles 
case which we use in Figs. 16-18, we find a 7\% reduction 
of charge fluctuations. This is a greater reduction than the
pure neutral resonance case. In our bubble model the decay 
particles are well contained and focused into our cut region.
 
Thus considering the foregoing analyses, we conclude that background
resonance effects will not to any significant degree affect our 
conclusions based on our simulations.
 
\begin{center}
\begin{tabular}{|c|r|r|r|r|}\hline
\multicolumn{5}{|c|}{Table II}\\ \hline                                 
particle & number & temp & width & $V_2$ \\ \hline
$\sigma$ & 69 & .700 & 1.80 & .04 \\ \hline
$\rho^0$ & 20 & .700 & 1.80 & .04 \\ \hline
$\sigma$ & 1246 & .335 & 1.80 & .07 \\ \hline
$\rho^0$ & 380 & .335 & 1.80 & .08 \\ \hline
\end{tabular}
\end{center}

\section{Summary and Conclusions}
 
From the early days of QGP theories we have shared\cite{workshop} the 
viewpoint of Leon Van Hove\cite{van83,van87}, that localized in           
phase space bubbles of QGP are more likely to be the origin of the 
hadronization products which originate from a QGP. This is especially
the case for a QGP  produced at high energy heavy ion colliders such 
as RHIC and LHC.
 
In a previous publication\cite{lind00} we have considered the case 
of one to at most a few separated QGP bubbles being produced.  This 
could conceivably occur, and be observed with sufficient statistics. 
We referred to multiple bubble formation in Ref. \cite{lind00}, 
which is the most likely high cross-section case.  However we did not 
calculate that case in Ref. \cite{lind00}. In this paper we 
conservatively calculated the multiple bubble case. The bubbles 
were embedded in HIJING, suitably modified for relevant elliptic 
flow effects.

We demonstrated that resonance effects will not significantly 
affect our conclusions.  The first test of our ideas will 
come from forthcoming STAR data at RHIC.  Therefore we made every 
reasonable effort to cast our predictions in a form which could 
be directly and quantitatively compared with the 
STAR data for 65GeV/n $Au$ colliding with 65 GeV/n $Au$. Obviously 
the methods and models we used could also be used for higher 
energy RHIC data, 
and eventually for LHC data.  It is expected that high statistics 
data on 100GeV/n $Au$ colliding with 100 GeV/n $Au$ will 
subsequently become available in the near future and allow a more 
critical comparison with these ideas.  We used a 2 fm radius bubble 
size based on HBT work\cite{adler2}. This work shows that above 
$p_t$ of 0.8 GeV/c the source size has a radius of about 2 fm. 
This is consistent with a 2 fm radius bubble size in Ref. \cite{dumi01} 
(one of their choices). We showed that a reasonable estimate of 
the outer shell of a ring of adjoining bubbles around the blast 
region, at central rapidity, when hadronization takes place, is 
located  at a radius of about 8 fm. The width of this outer shell
is roughly $\pm$ 2 fm about the 8 fm radius. We added a reasonable    
estimate of longitudinal expansion to the bubbles in order to take 
account of the expected Landau effect, but 
comparisons with the data can determine this value.

The lower bound in $p_t$ is chosen to maintain the space momentum 
correlation corresponding to the 2 fm source size, which is consistent 
with the HBT work and enhances our signal. The upper bound in $p_t$ is 
chosen to avoid the contamination from the hard physics region 
above 2 GeV/c.  However at times we also raised the lower bound 
of the $p_t$ cut to 1.2 GeV/c, for various reasons as explained 
in the text. Since we estimated the $\Phi$ angular range to which 
each bubble contributes is about 30$^\circ$, we used an average of 13 
adjoining bubbles around a ring  in the central rapidity region.
This fills the $\Phi$ coverage. Our standard cut in $\eta$ was 
$-0.75$ to $+ 0.75$ to cover the central region dominated by our bubble 
ring hadronization products. Each bubble is considered to be a 
localized gluon dominated, uncharged color singlet system. Thus 
when the bubble hadronizes the total charge coming from the bubble 
was set equal to zero.
 
We have demonstrated by simulations signals for the QGP bubbles.
Below we list signals  which had a statistical significance of 
$5\sigma$ or more:

\begin{itemize}

\item{The} angular correlation of charged particles as a function 
of $\Delta\Phi$, and also opening angles in the 0-60$^\circ$ 
$\Delta\Phi$ range, where our bubbles contribute. 
 
\item{Predicted} suppression of electric-charge fluctuations.
 
\item{For} correlation vs $\Delta\Phi$: Fig. 2 compares modified 
HIJING and modified HIJING with the addition of bubbles, which results 
in an $8\sigma$ difference effect (see text) for our 0-60$^\circ$ 
cut where the effect of bubbles is expected to occur. This cut is 
used on all angular distributions. To remove this difference
requires the very conservative extreme measure of doubling the 
number of jets (as shown in Fig. 3 and text), when there are 
arguments that jet quenching occurs rather than enhancement.
 
\item{In} Fig. 10 we see that when we compare the $\Delta\Phi$ correlations 
for like and unlike charge particles there is a $5\sigma$ difference 
in the case of the bubbles plus modified HIJING, and modified HIJING
plus double jets. This is caused by charge suppression due to bubbles.
 
\item{We} pointed out\cite{lind01} that earlier work\cite{jeon,asakawa} 
on charge fluctuation suppression overlooked color charge effects,
and kinematic mixing effects, which could drastically reduce, or 
even more or less eliminate the calculated suppression. In our 
localized bubble scenario these effects are negligible. In our 
analysis the predicted observed charge suppression in the 65 GeV/n 
$Au$ on 65 GeV/n $Au$ simulations will range from about 5\% for the 
Ref. \cite{dumi01} based value of about 3.25 GeV per bubble going into 
charged particles, to about 27\% for about 10 GeV per bubble going 
into charged particles. In the case of 10 GeV per bubble going 
into charged particles, we find the difference in the $\Delta\Phi$ 
correlation between our modified HIJING in Fig. 2 and modified 
HIJING plus bubbles is huge (see Fig. 13 and text), and well 
over $20\sigma$. Also in Fig. 14 using the same modified HIJING 
plus bubbles model the difference in the $\Delta\Phi$ correlation 
between like and unlike charge correlation's is huge, and well 
over $20\sigma$.
\end{itemize}

\begin{figure}
\begin{center}               
\mbox{
   \epsfysize 4.0in
   \epsfbox{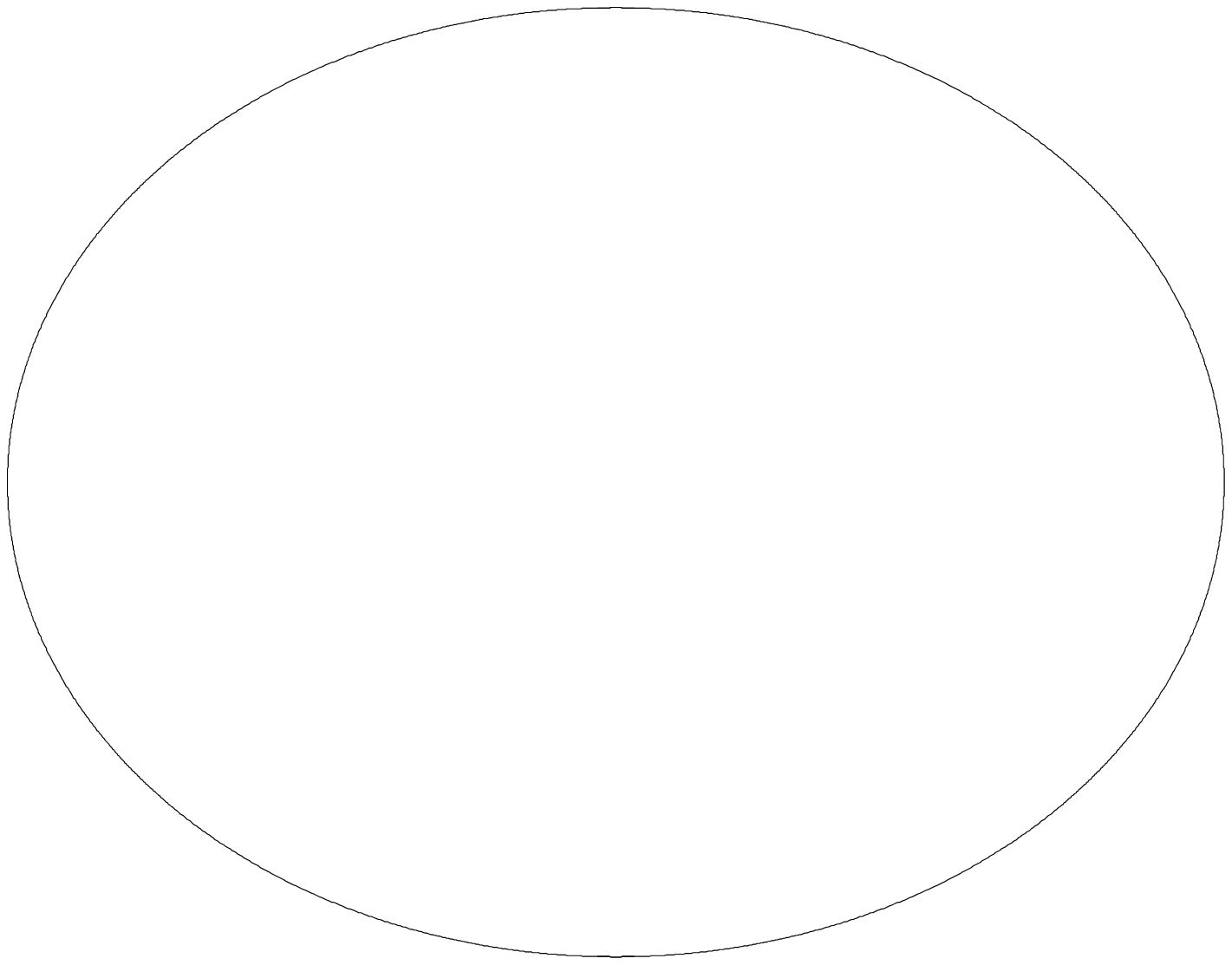}}
\end{center}
\caption{is a simplified attempt by using a classical (not quantum 
mechanical) sketch to illustrate our bubble geometry.  We have used 
2 fm radius spherical bubbles elongated in the longitudinal direction 
by the Landau effect.  The best value of the longitudinal expansion 
will be determined by the data analysis. We have shown a section of 
a bubble parallel to its direction of motion which illustrates the 
longitudinal effect, when looking at the figure in the 
horizontal direction. A ring of 13 of these adjoining originally 
spherical bubbles was placed near central rapidity $\eta\approx 0$ 
at an approximate radius of 8 fm and the longitudinal elongation 
developed as they moved before hadronization. The description of 
our procedure of this complex process is given in the text.}
\label{fig1}
\end{figure}
 
\begin{figure}
\begin{center}               
\mbox{
   \epsfysize 4.4in
   \epsfbox{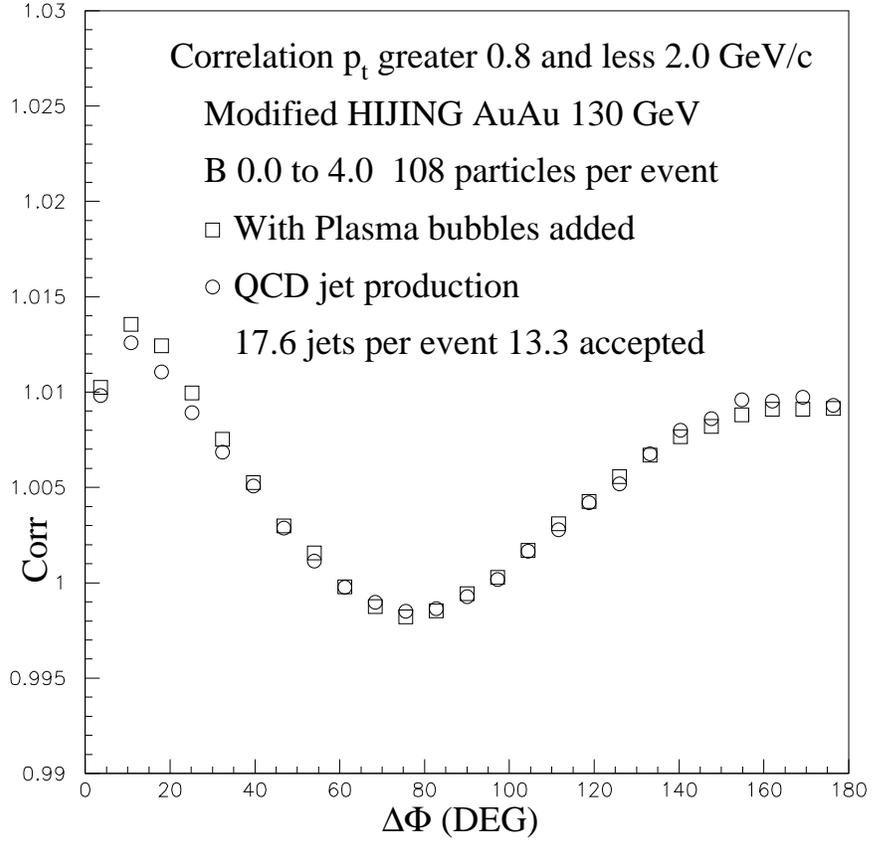}}
\end{center}
\caption{The $\Delta\Phi$ correlation of charged particles which lie
between $p_t$ (transverse momentum) 0.8 to 2.0 GeV/c for 
two different models based on HIJING (see text). The circles
are HIJING (with elliptic flow) plus jets and the squares
are HIJING (with elliptic flow) plus plasma bubbles (see text). 
Also absolute $\eta$ (pseudo-rapidity) $< 0.75 $ is required.}
\label{fig2}
\end{figure}
 
\begin{figure}
\begin{center}
\mbox{
   \epsfysize 4.4in
   \epsfbox{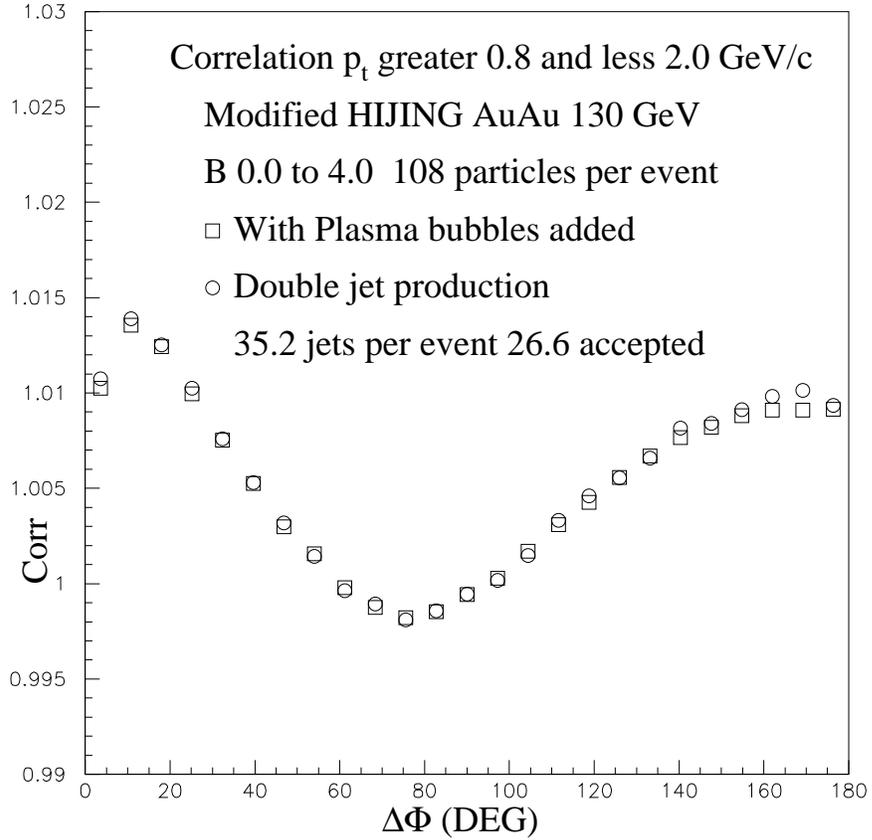}}
\end{center}
\caption{The $\Delta\Phi$ correlation of charged particles 
($0.8<p_t<2.0$ GeV/c, $|\eta| < 0.75$) for two different models 
based on HIJING. The circles are HIJING plus double the number of 
expected jets and the squares are HIJING plus plasma bubbles. In 
calculating $\chi ^2$ values for differences (in text) the data 
in every figure were always cut for an angular range from 0 to 
60$^\circ$, since that is where we expect the bubble effects to occur.}
\label{fig3}
\end{figure}
 
\begin{figure}
\begin{center}
\mbox{
   \epsfysize 4.4in
   \epsfbox{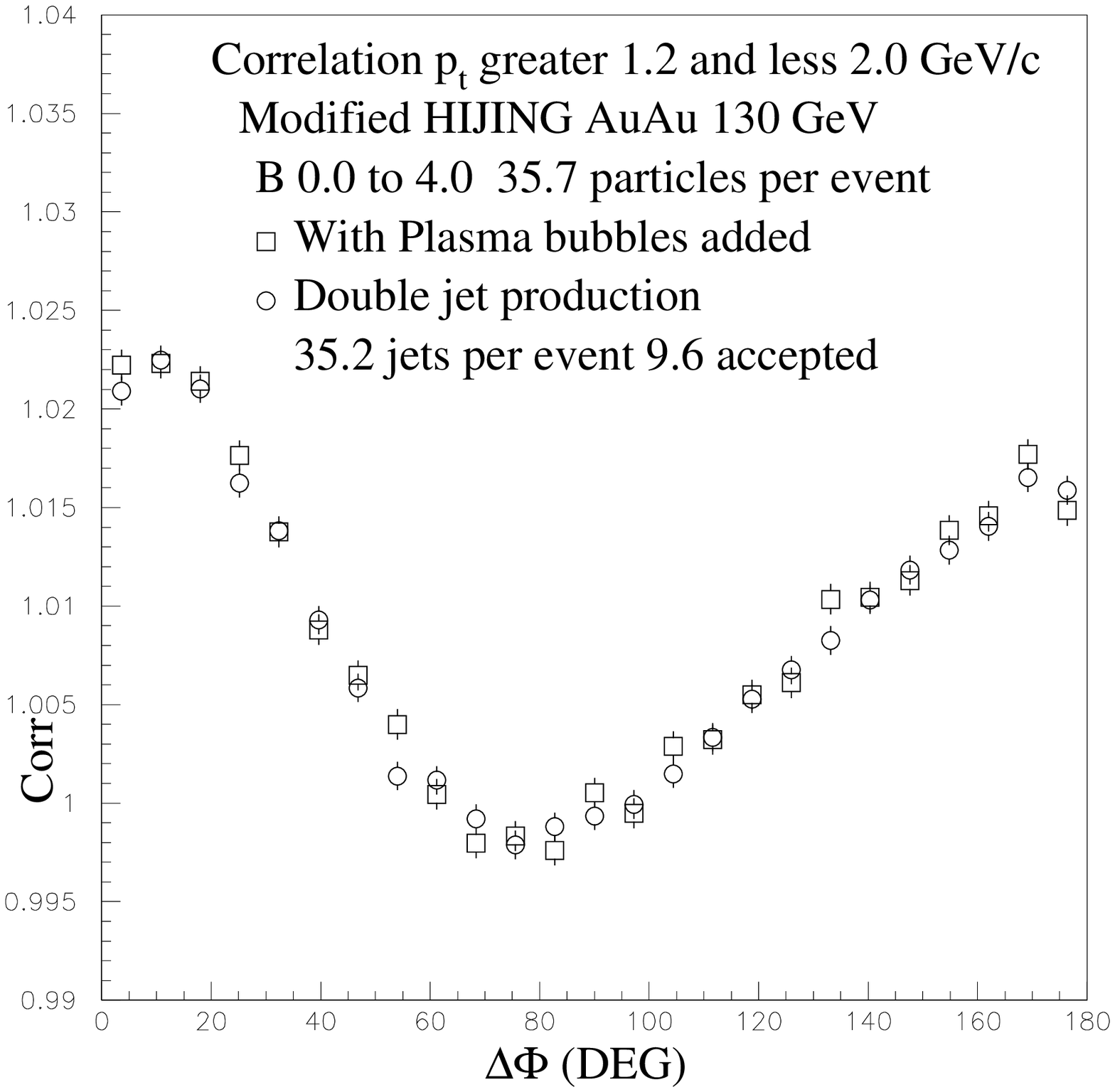}}
\end{center}
\vspace{2pt}
\caption{The $\Delta\Phi$ correlation of charged particles 
($1.2<p_t<2.0$ GeV/c, $|\eta| < 0.75$) for the same models as Fig. 3.}
\label{fig4}
\end{figure}
 
\begin{figure}
\begin{center}
\mbox{
   \epsfysize 4.4in
   \epsfbox{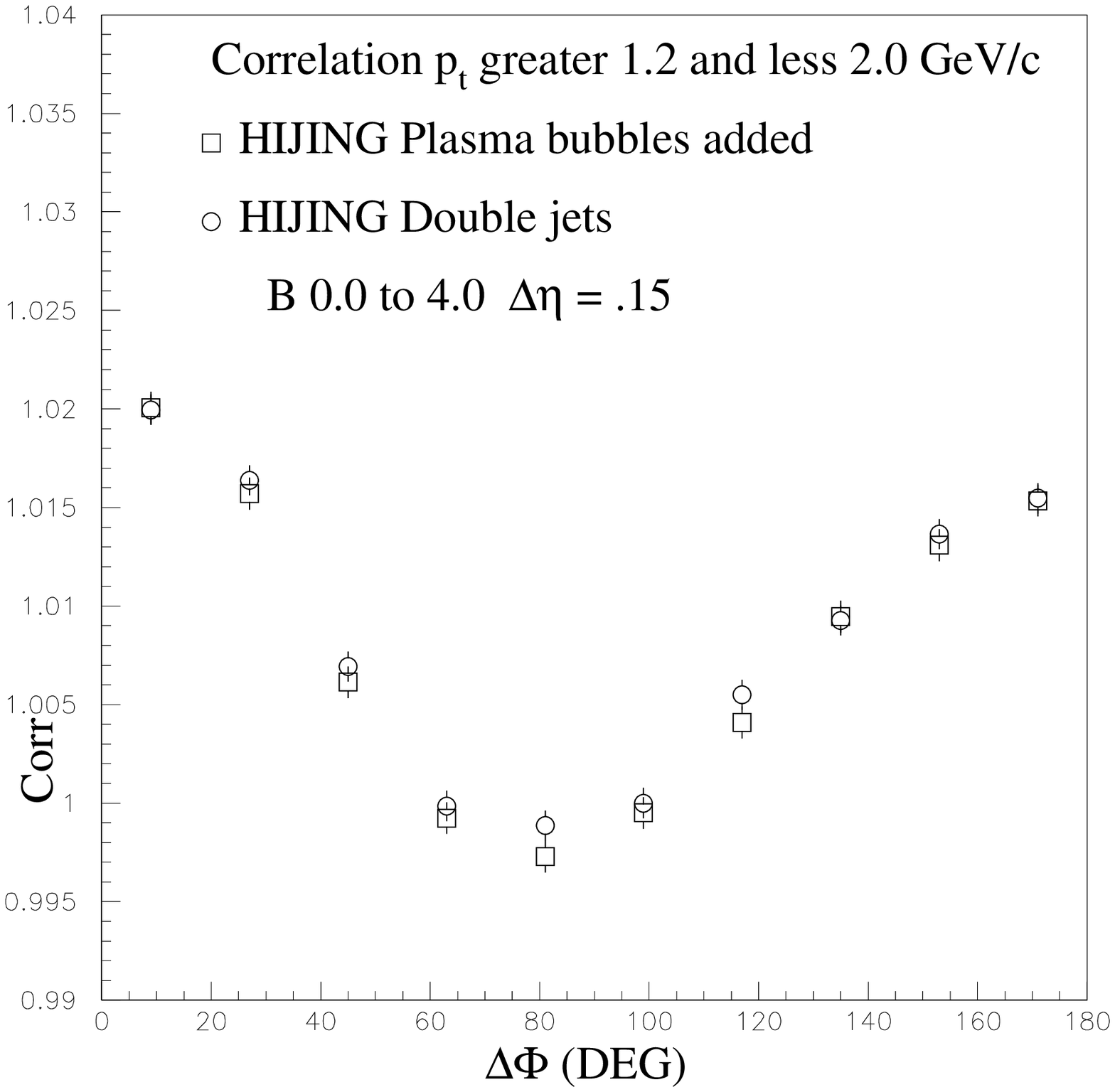}}
\end{center}
\vspace{2pt}
\caption{The $\Delta\Phi$ correlation of charged particles
($1.2<p_t<2.0$ GeV/c, $|\eta| < 0.75$) where the difference between 
the $\eta$ of the two charged particle is between 
$0.0< |\Delta\eta | < 0.3$ for the same models as Fig. 3 .}
\label{fig5}
\end{figure}
 
\begin{figure}
\begin{center}
\mbox{
   \epsfysize 4.4in
   \epsfbox{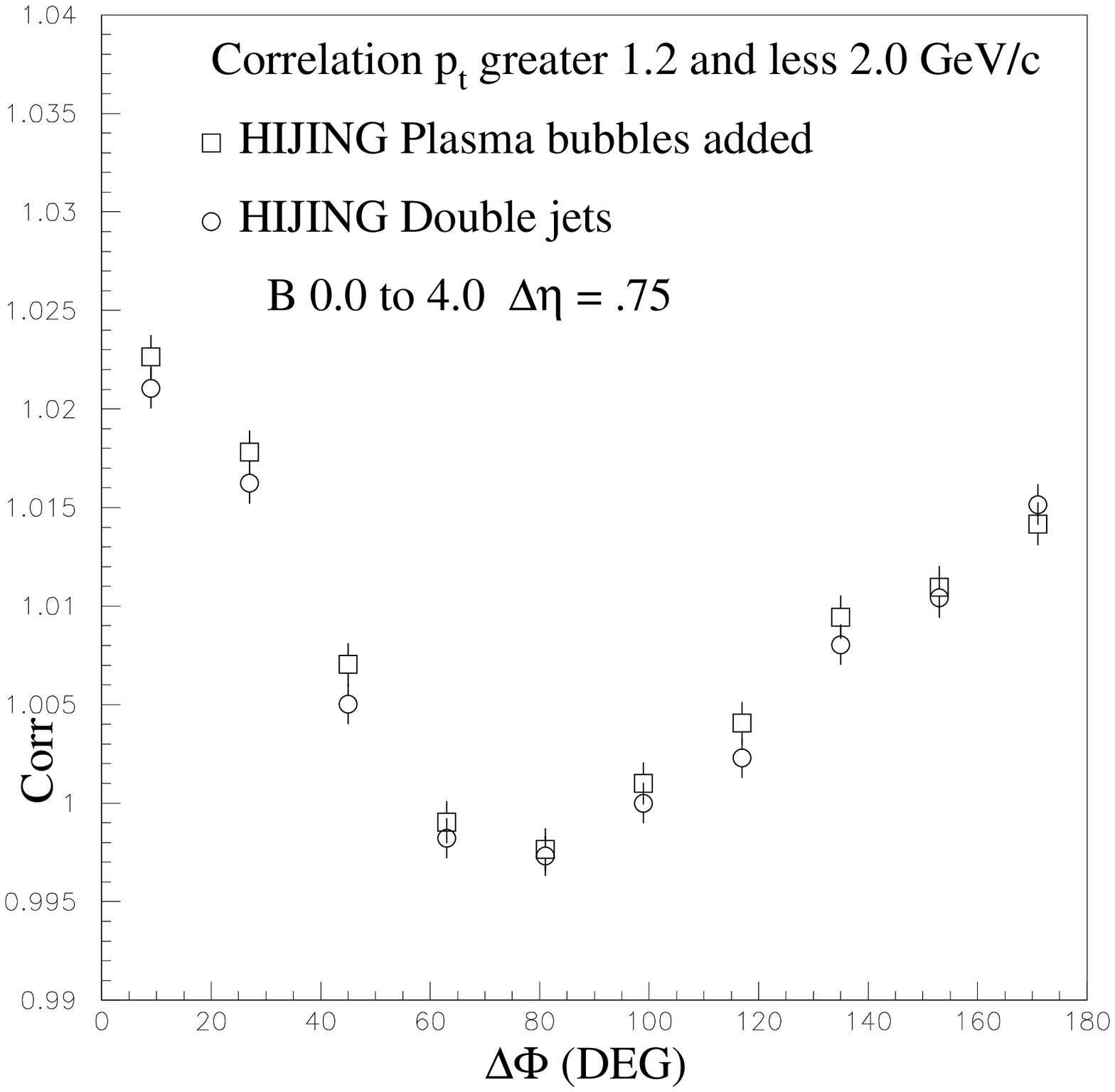}}
\end{center}
\vspace{2pt}
\caption{The $\Delta\Phi$ correlation of charged particles
($1.2<p_t<2.0$ GeV/c, $|\eta| < 0.75$), where the difference between 
the $\eta$ of the two charged particles is between 
$0.6< |\Delta\eta | < 0.9$ for the same models as Fig. 3 .}
\label{fig6}
\end{figure}
 
\begin{figure}
\begin{center}
\mbox{
   \epsfysize 4.4in
   \epsfbox{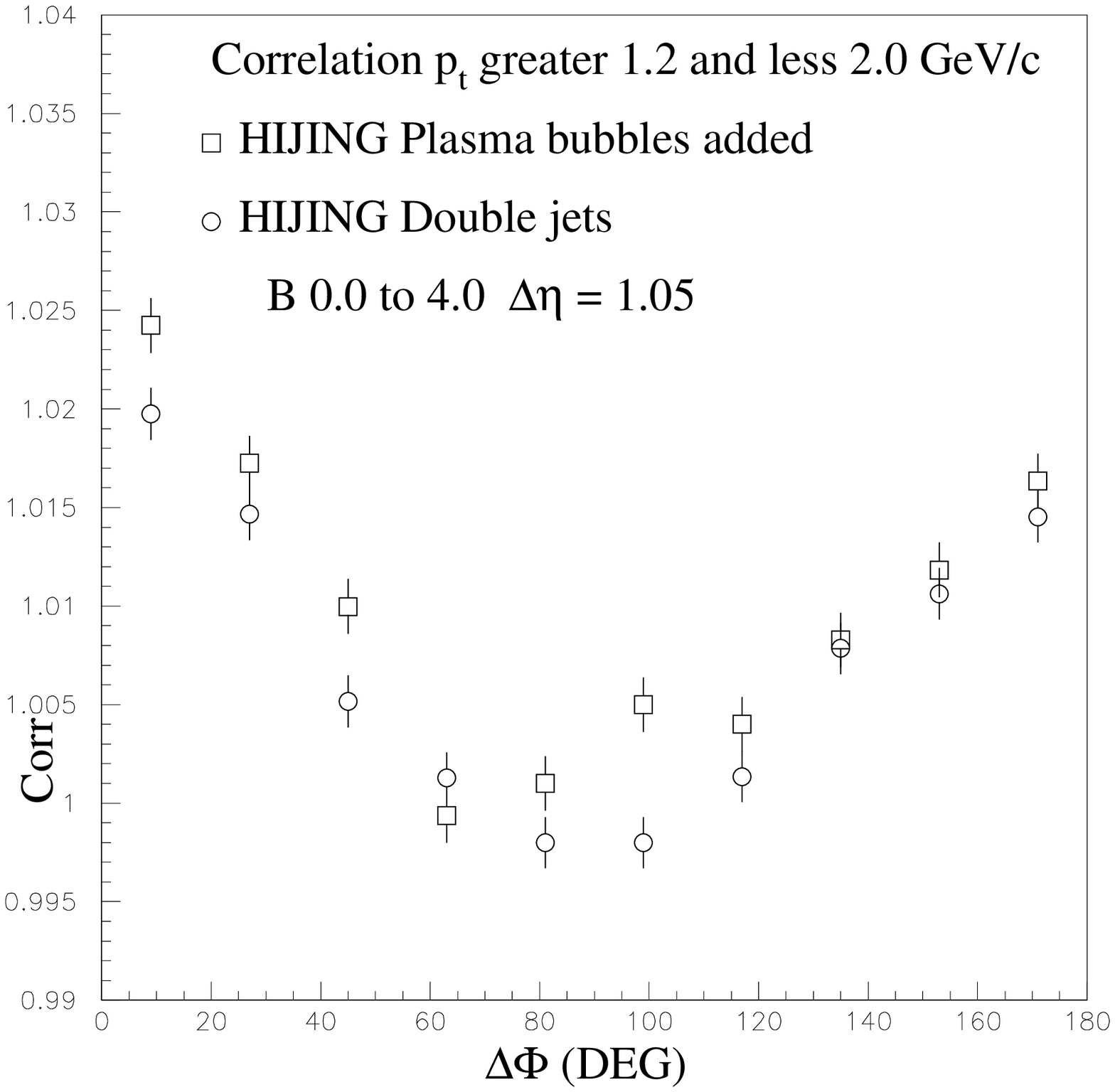}}
\end{center}
\vspace{2pt}
\caption{The $\Delta\Phi$ correlation of charged particles
($1.2<p_t<2.0$ GeV/c, $|\eta| < 0.75$), where the difference between 
the $\eta$ of the two charged particles is between 
$0.9< |\Delta \eta | < 1.2$ for the same models as Fig. 3.}
\label{fig7}
\end{figure}
 
\begin{figure}
\begin{center}
\mbox{
   \epsfysize 4.4in
   \epsfbox{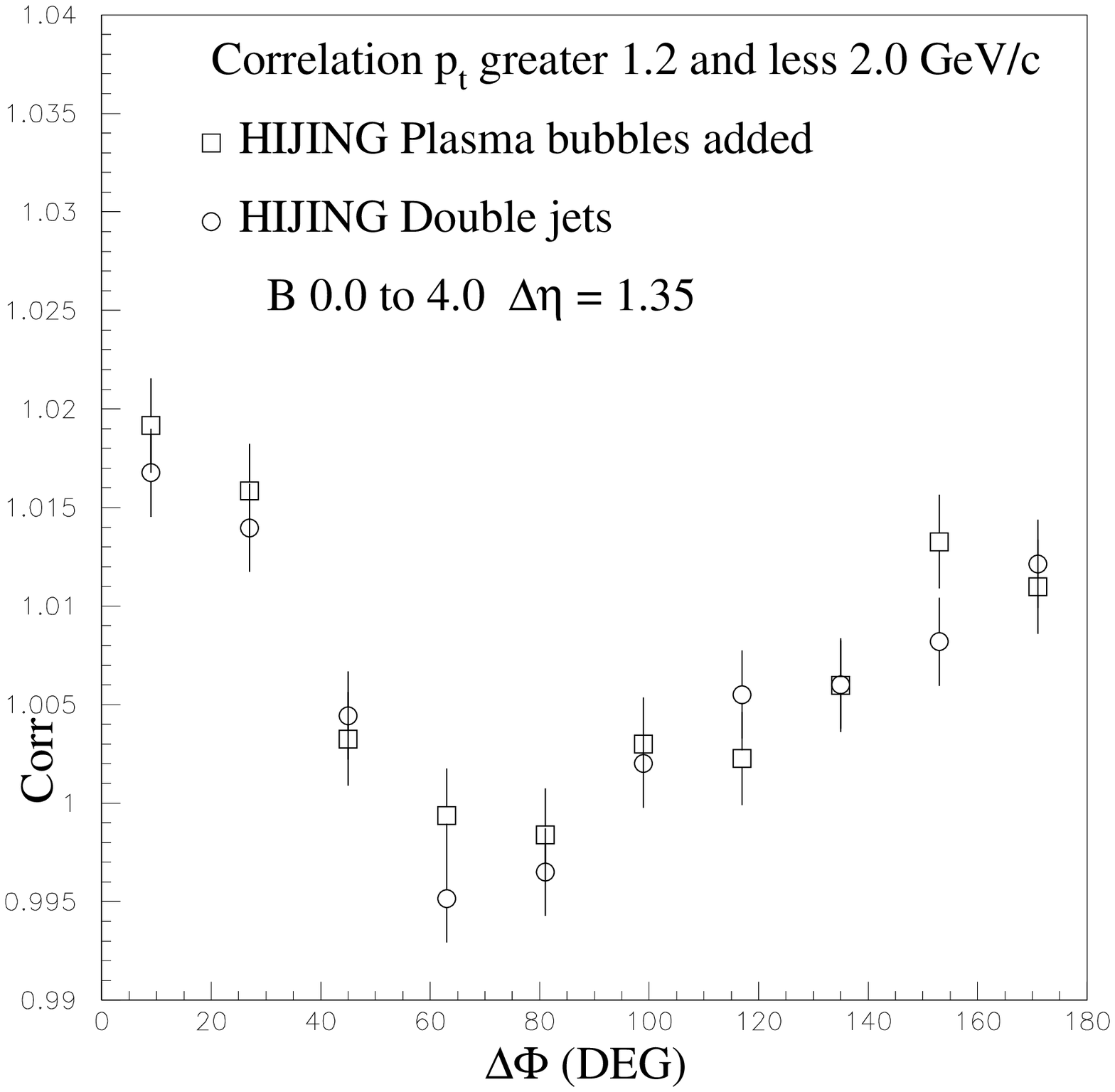}}
\end{center}
\vspace{2pt}
\caption{The $\Delta\Phi$ correlation of charged particles
($1.2<p_t<2.0$ GeV/c, $|\eta| < 0.75$), where the difference between 
the $\eta$ of the two charged particles is between 
$1.2< |\Delta \eta | < 1.5$ for the same models as Fig. 3.}
\label{fig8}
\end{figure}
 
\begin{figure}
\begin{center}
\mbox{
   \epsfysize 4.4in
   \epsfbox{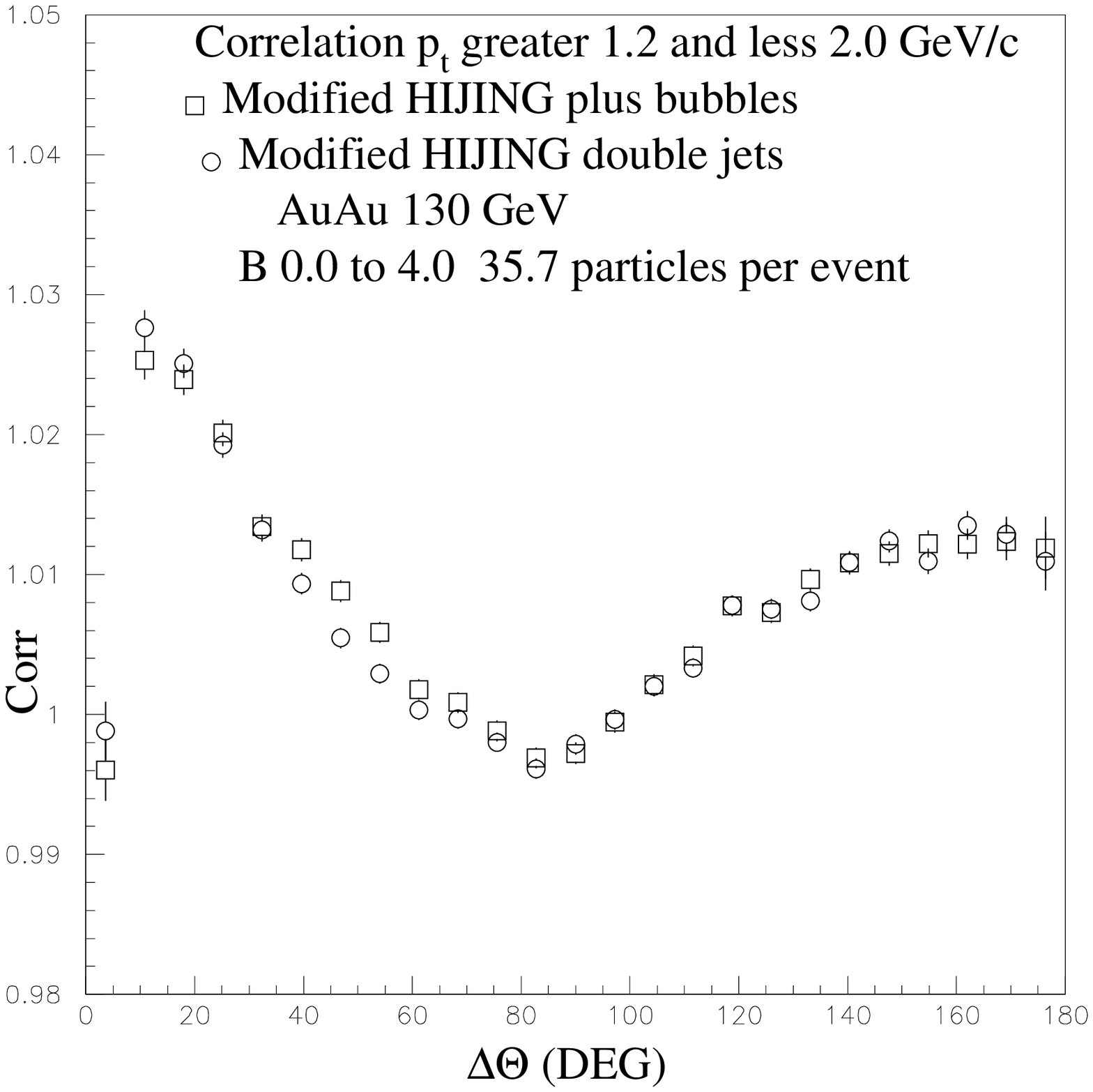}}
\end{center}
\vspace{2pt}
\caption{The $\Delta\Theta$ correlation of charged particles
(where $\Theta$ is the opening angle and ($1.2<p_t<2.0$ GeV/c, 
$|\eta| < 0.75$ are cuts) for the same models as Fig. 3.}
\label{fig9}
\end{figure}
 
\begin{figure}
\begin{center}
\mbox{
   \epsfysize 4.4in
   \epsfbox{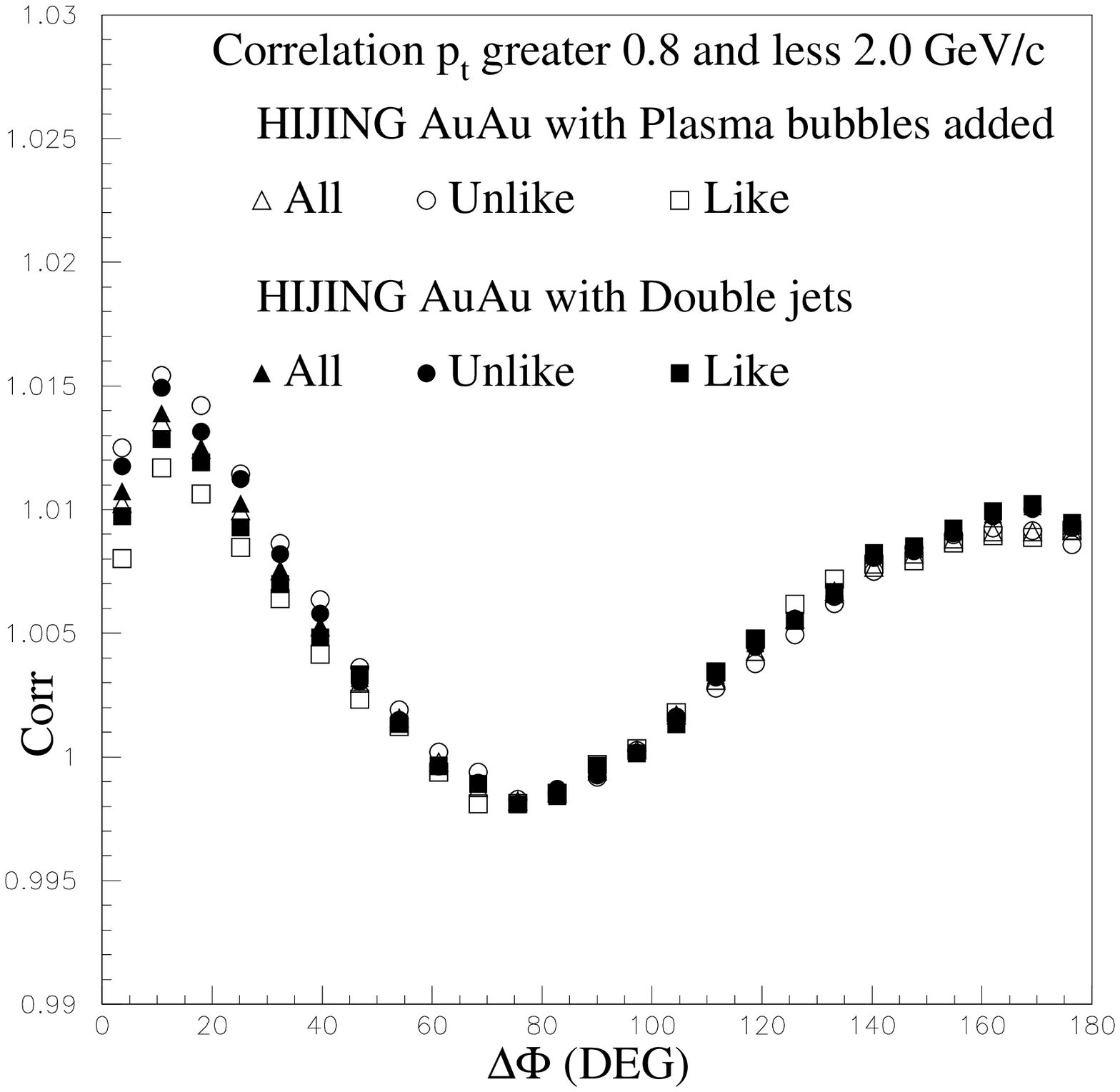}}
\end{center}
\vspace{2pt}
\caption{The $\Delta\Phi$ correlation of charged particles
($0.8<p_t<2.0$ GeV/c, $|\eta| < 0.75$) for the same models as Fig. 3. 
The open triangles are the same as the squares and 
the solid triangles are the same as the circles of Fig. 2. 
The circles are the unlike sign particles and the square are the 
like sign particles.}
\label{fig10}
\end{figure}

\begin{figure}
\begin{center}
\mbox{
   \epsfysize 4.4in
   \epsfbox{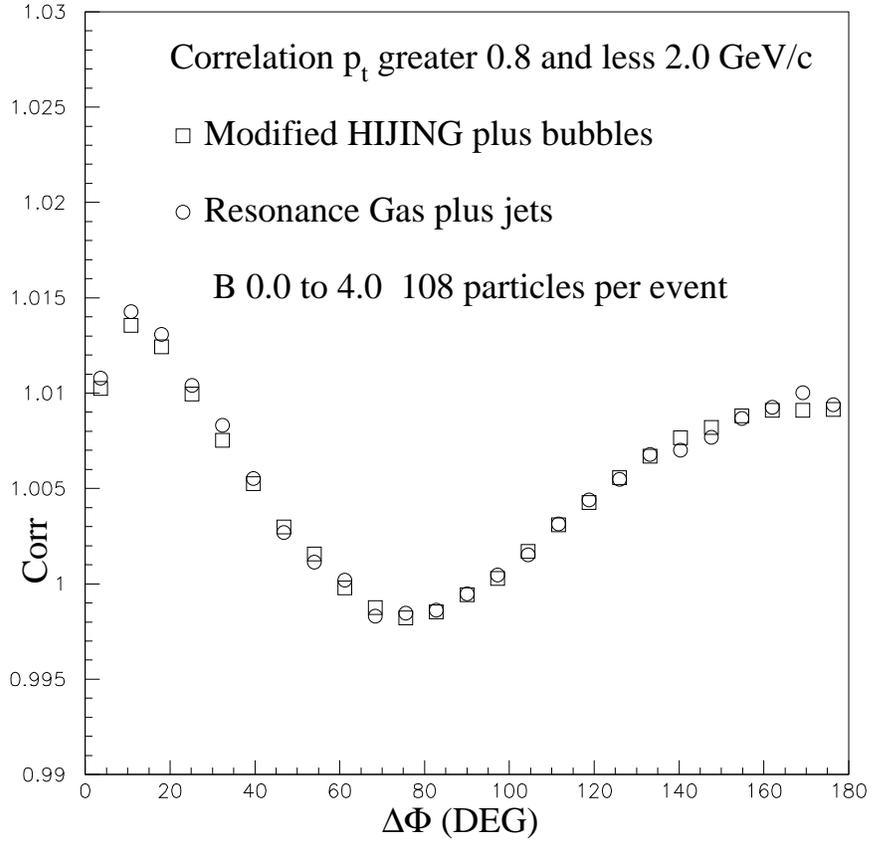}}
\end{center}
\caption{The $\Delta\Phi$ correlation of charged particles
($0.8<p_t<2.0$ GeV/c, $|\eta| < 0.75$) for two different models. 
The squares are HIJING plus plasma bubbles which are the same
as in Fig. 3. The circles are the resonance model as described in the 
text.}
\label{fig11}
\end{figure}
 
\begin{figure}
\begin{center}
\mbox{
   \epsfysize 4.4in
   \epsfbox{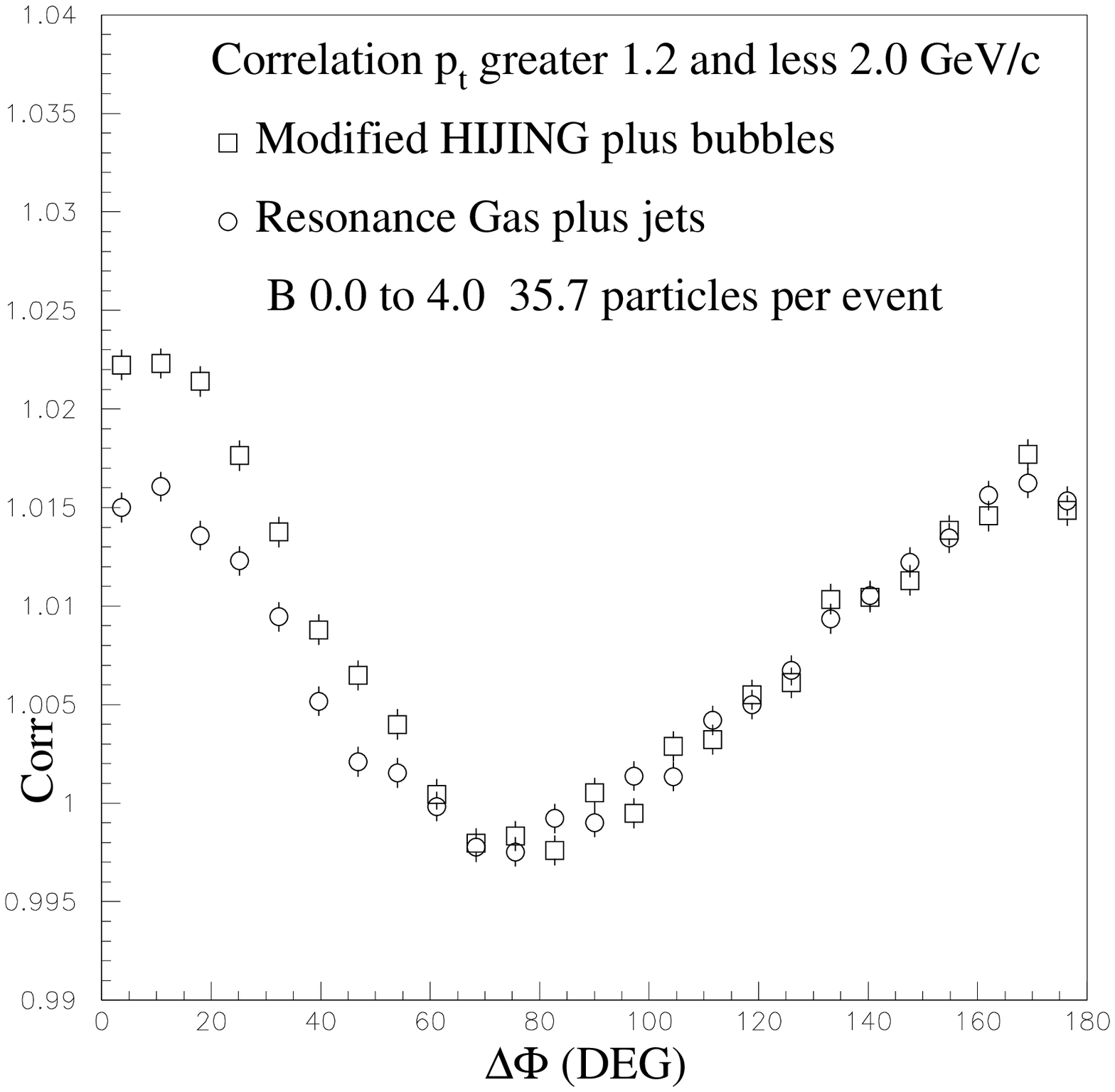}}
\end{center}
\caption{The $\Delta\Phi$ correlation of charged particles
($1.2<p_t<2.0$ GeV/c, $|\eta| < 0.75$) for the same models as 
Fig. 11.}
\label{fig12}
\end{figure}
 
\begin{figure}
\begin{center}
\mbox{
   \epsfysize 4.4in
   \epsfbox{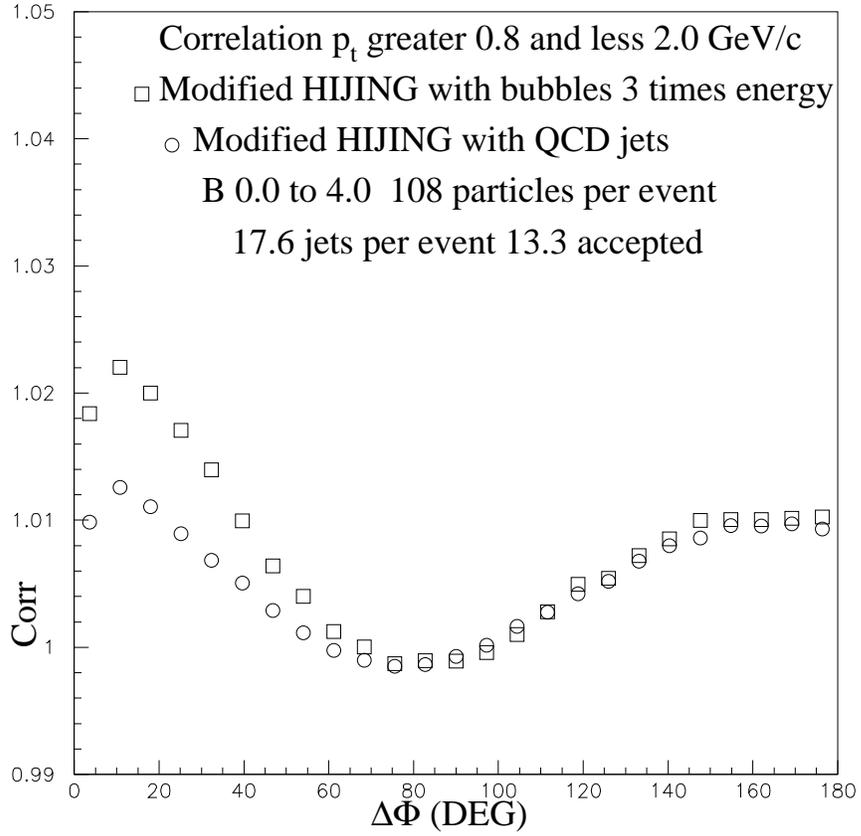}}
\end{center}
\caption{The $\Delta\Phi$ correlation of charged particles
($0.8<p_t<2.0$ GeV/c, $|\eta| < 0.75$) for two different models
based on HIJING. The circles are HIJING plus the normal number of expected 
jets and the squares are HIJING plus plasma bubbles which
have 3 times the energy of the plasma bubbles used in Fig. 2.}
\label{fig13}
\end{figure}
 
\begin{figure}
\begin{center}
\mbox{
   \epsfysize 4.4in
   \epsfbox{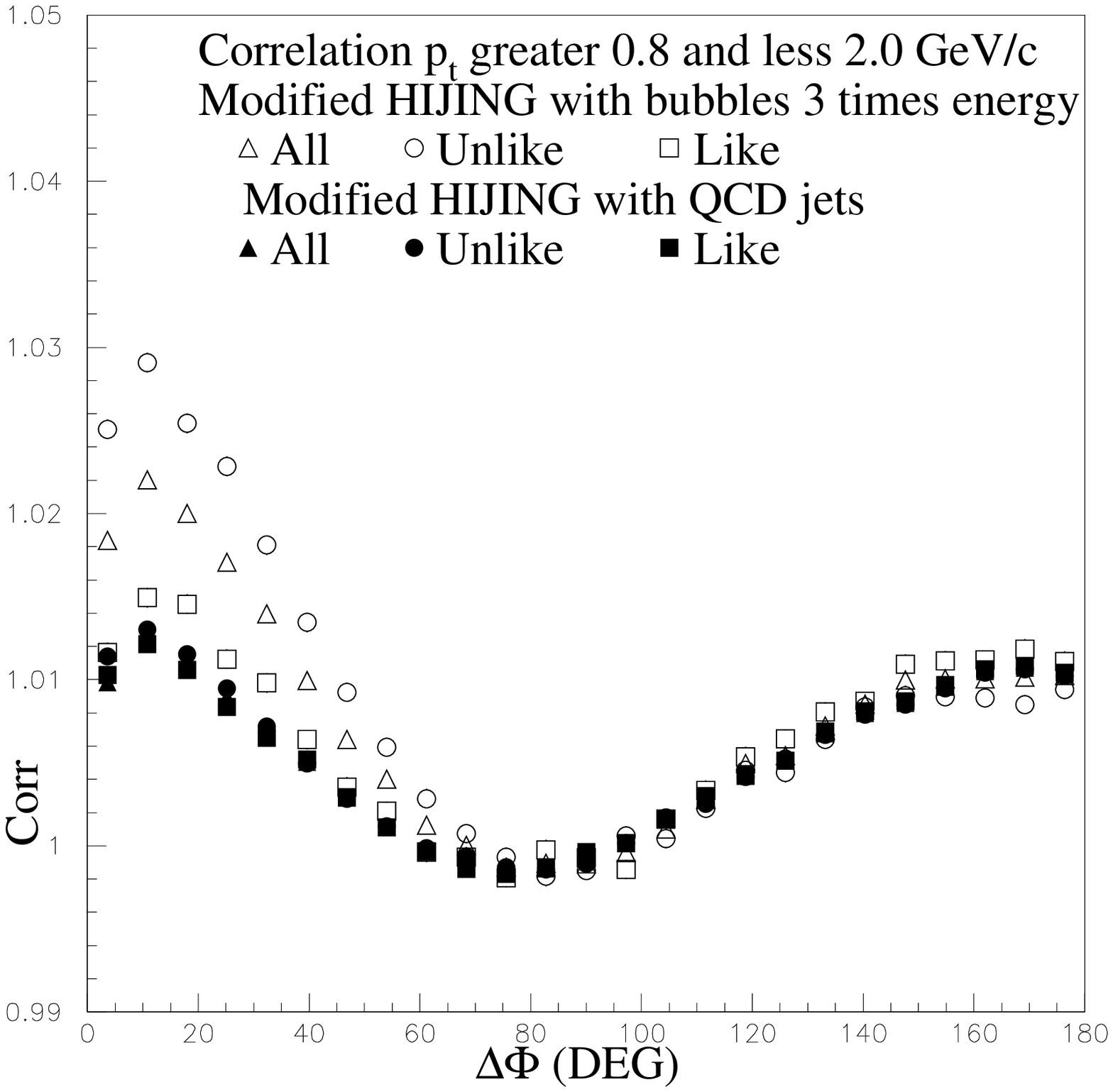}}
\end{center}
\caption{The $\Delta\Phi$ correlation of charged particles
($0.8<p_t<2.0$ GeV/c, $|\eta| < 0.75$) for the same models as Fig. 13. 
The open triangles are the same as the squares and the 
solid triangles are the same as the circles of Fig. 13. The circles 
are the unlike sign particles and the squares are the like sign particles.}
\label{fig14}
\end{figure}
 
\begin{figure}
\begin{center}
\mbox{
   \epsfysize 4.4in
   \epsfbox{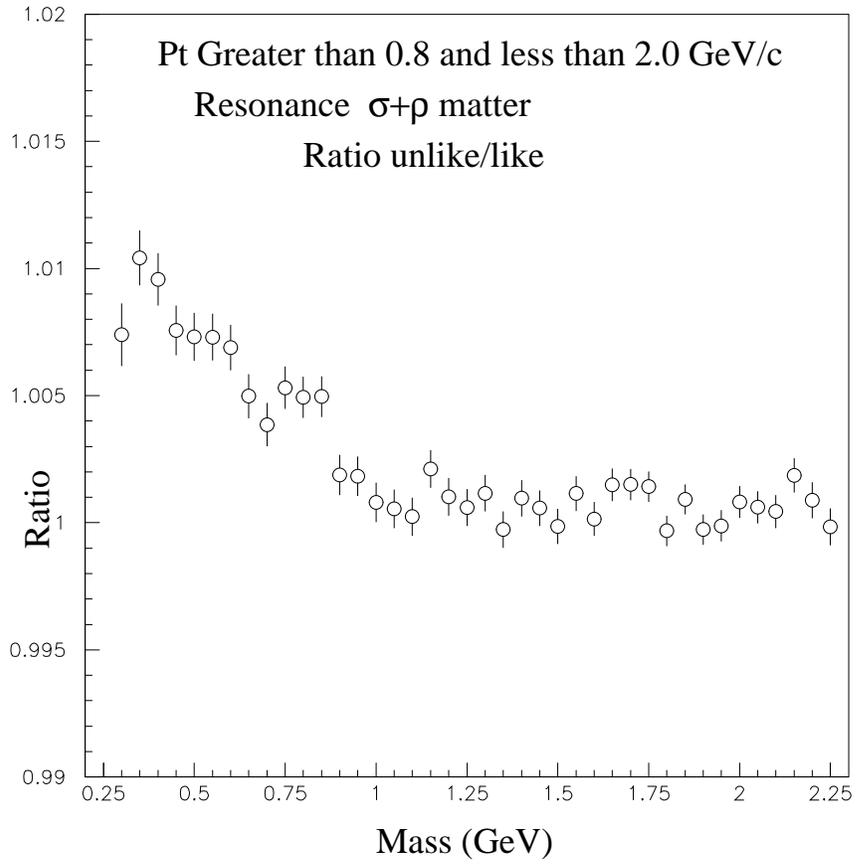}}
\end{center}
\caption{The ratio unlike to like of the effective mass spectrum
($0.8<p_t<2.0$ GeV/c, $|\eta| < 0.75$) for a pure neutral 
resonance model (see text).}
\label{fig15}
\end{figure}

\begin{figure}
\begin{center}
\mbox{
   \epsfysize 4.4in
   \epsfbox{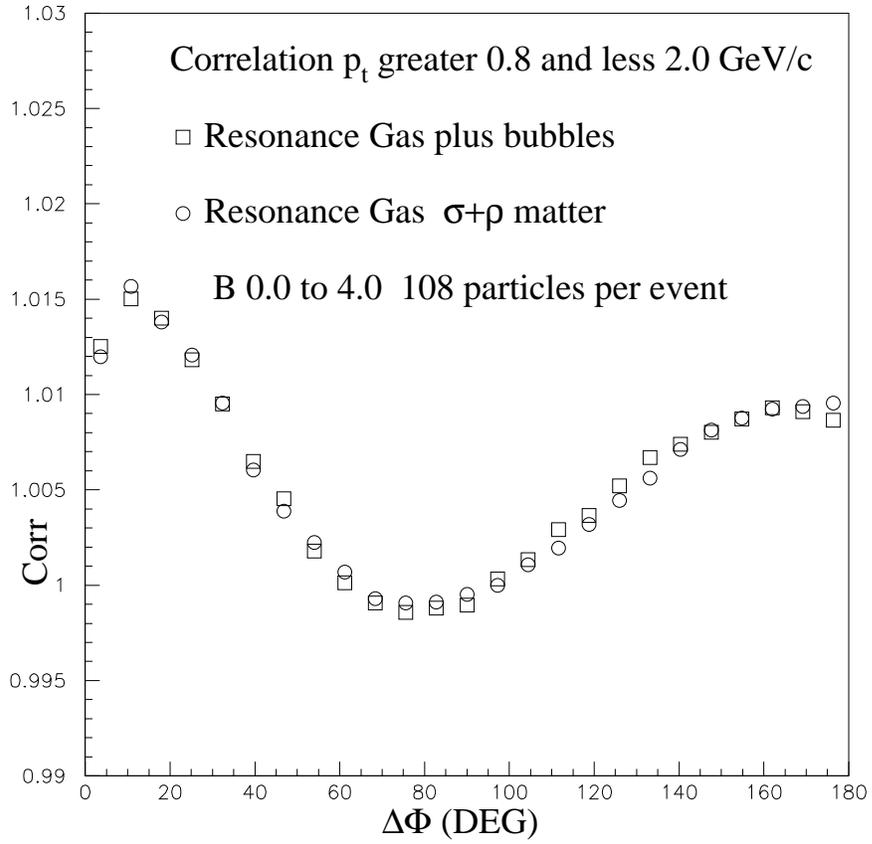}}
\end{center}
\caption{The $\Delta\Phi$ correlation of charged particles
($0.8<p_t<2.0$ GeV/c, $|\eta| < 0.75$) for two different models.
One being the resonance gas model plus plasma bubbles (squares), 
and the other being the pure neutral resonance of Fig. 15 (circles).}
\label{fig16}
\end{figure}
 
\begin{figure}
\begin{center}
\mbox{
   \epsfysize 4.4in
   \epsfbox{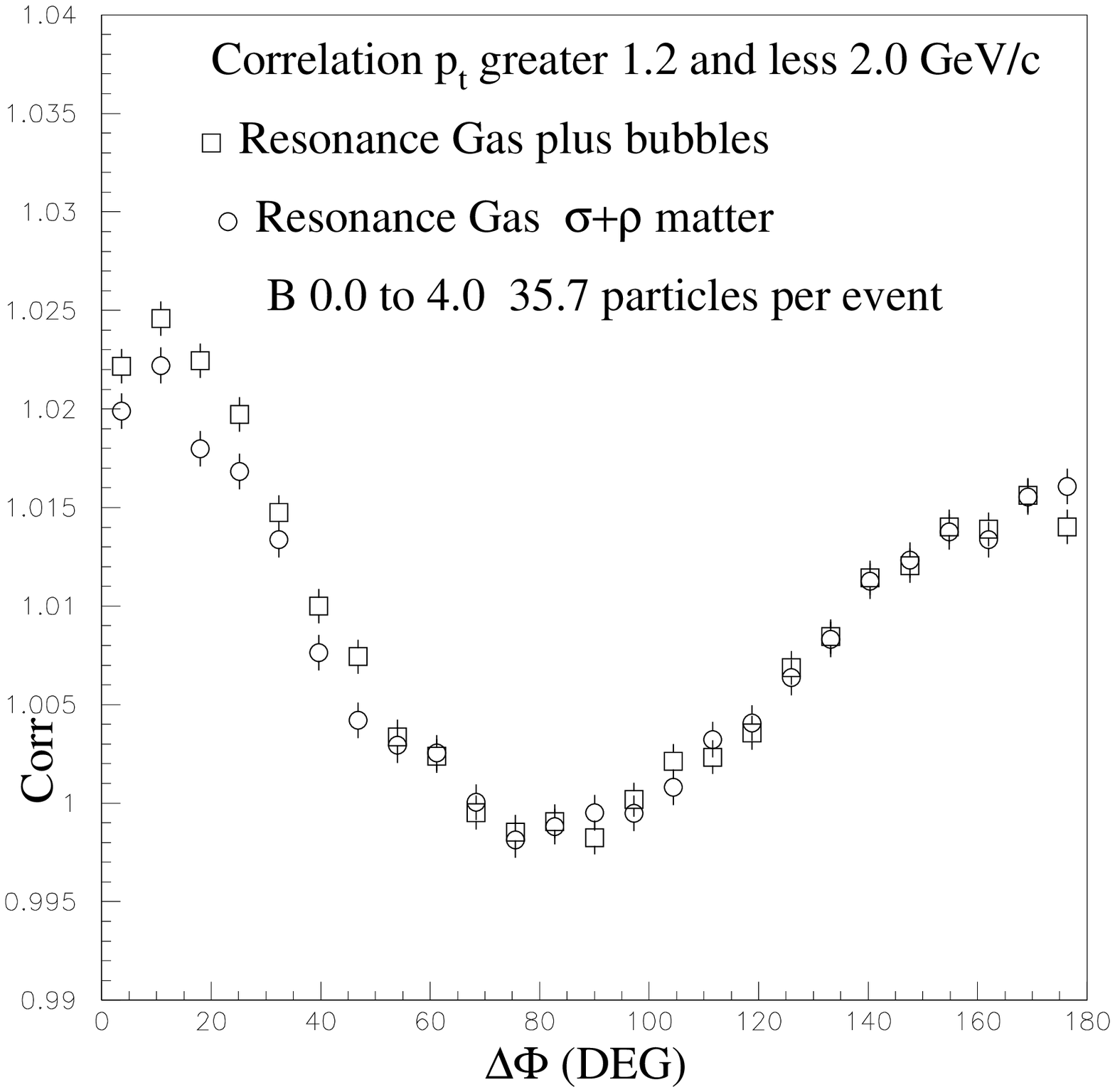}}
\end{center}
\caption{The $\Delta\Phi$ correlation of charged particles
($1.2<p_t<2.0$ GeV/c, $|\eta| < 0.75$) for the same models as Fig. 16.}
\label{fig17}
\end{figure}
 
\begin{figure}
\begin{center}
\mbox{
   \epsfysize 4.4in
   \epsfbox{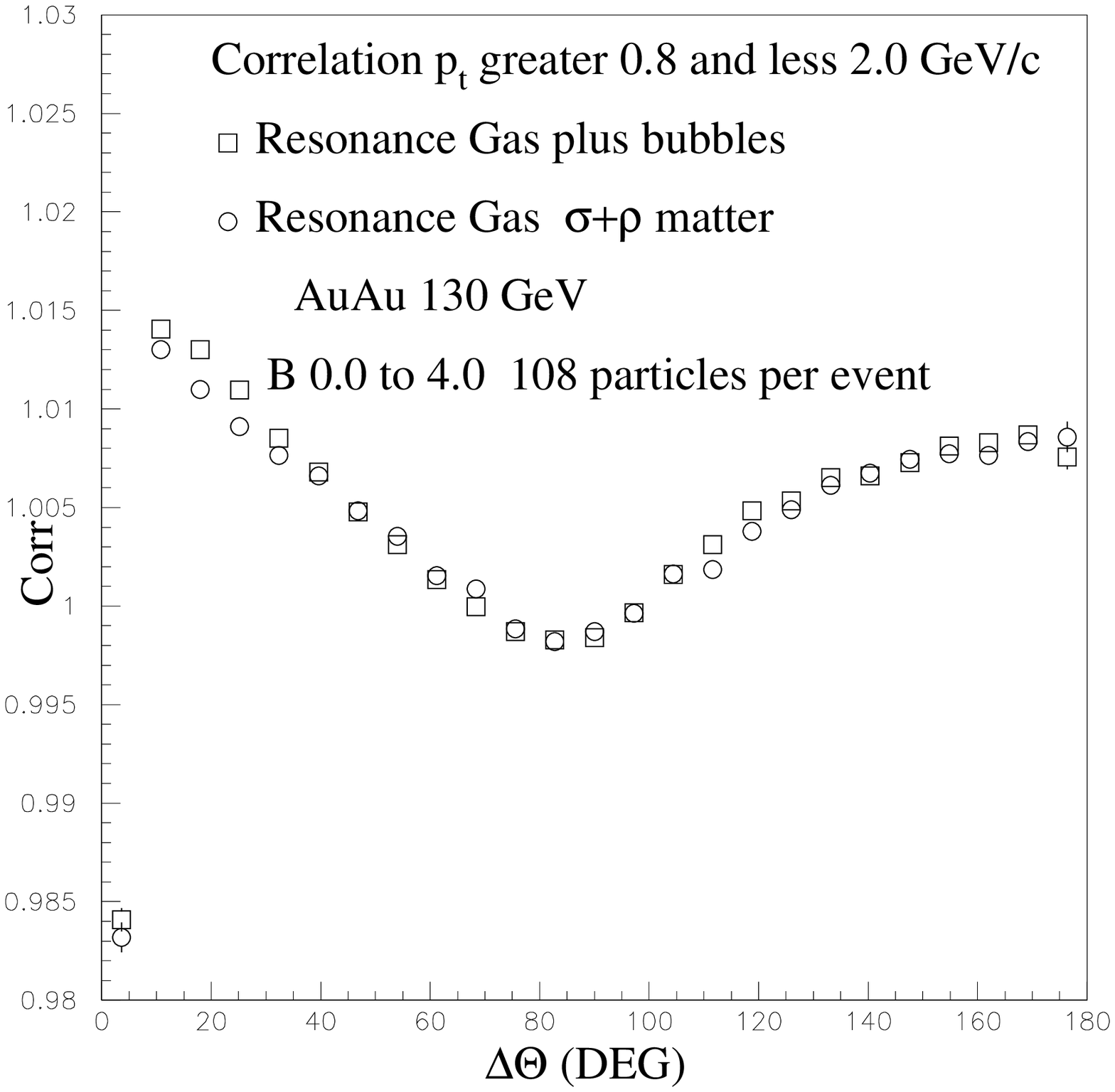}}
\end{center}
\caption{The $\Delta\Theta$ correlation of charged particles 
(where $\Theta$ is the opening angle and $1.2<p_t<2.0$ GeV/c, 
$|\eta| < 0.75$ are cuts) for the same models as Fig. 16.}
\label{fig18}
\end{figure}
\end{document}